\documentclass[usenatbib]{mn2e}

\usepackage{astrojournals}
\usepackage{graphicx} 
\usepackage{amssymb,amsmath}
\usepackage{times}
\usepackage{url}

\begin{document}

\bibliographystyle{mn2e}

\title [SIDM Halo Shapes]{Cosmological Simulations with Self-Interacting Dark Matter II: Halo Shapes vs. Observations}

\author[Peter et al.] 
{
Annika H. G. Peter,$^{1}$\thanks{E-mail: annika.peter@uci.edu} 
Miguel Rocha,$^1$\thanks{E-mail: rocham@uci.edu}
James S. Bullock,$^1$\thanks{E-mail: bullock@uci.edu}
and Manoj Kaplinghat$^1$\thanks{E-mail: mkapling@uci.edu}\\
$^1$Center for Cosmology, Department of Physics and Astronomy, University of California, Irvine, CA
92697-4575, USA\\
}

\maketitle
\date{\today}

\begin{abstract} 
If dark matter has a large self-interaction scattering cross section, then interactions among dark-matter particles will drive galaxy and cluster halos to become spherical in their centers.   Work in the past has used this effect to rule out velocity-independent, elastic cross sections larger than
$\sigma/m \simeq 0.02\hbox{ cm}^2/\hbox{g}$ based on comparisons to the shapes of galaxy cluster lensing potentials and X-ray isophotes.
  In this paper, we use cosmological simulations to show that these constraints were off by more than an order of magnitude because (a) they did not properly account for the fact that the observed ellipticity gets contributions from the triaxial mass distribution outside the core set by scatterings, (b) the scatter in axis ratios is large and (c) the core region retains more of its triaxial nature than estimated before. Including these effects properly shows that the same observations now allow dark matter self-interaction cross sections at least as large as $\sigma/m = 0.1\hbox{ cm}^2/\hbox{g}$.  We show that constraints on self-interacting dark matter from strong-lensing clusters are likely to improve significantly in the near future, but possibly more via central densities and core sizes than halo shapes.  
\end{abstract}

\begin{keywords}
galaxies: halos --- methods: numerical                            
\end{keywords} 

\section{Introduction}
The nature of dark matter is one of the most compelling mysteries of our time.  On large scales, the behavior of dark matter is consistent with what cosmologists of yore called ``dust'' \citep[e.g.,][]{tolman1934}, meaning its behavior is consistent with being collisionless and non-relativistic (``cold'') 
for the vast majority of the Universe's history \citep{breid2010}.  This consistency has been of great interest to the particle-physics community because the most popular candidate for dark matter, the supersymmetric neutralino, displays exactly this behavior \citep{steigman1985,griest1988b,jungman1996}.  While the supersymmetric neutralino paradigm is attractive in many ways, there are two outstanding problems with it.  First, astroparticle searches have yet to turn up evidence for the existence of the neutralino, 
though searches are rapidly increasing their sensitivity to interesting neutralino parameter space \citep{geringer-sameth2011,ackermann2011a,cotta2011,fox2011b,bertone2012,atlas2012,koay2012,baudis2012,baer2012,xenon2012}.  Second, there are predictions for the structure of dark-matter halos that have not been observationally verified at a quantitative level \citep{dubinski1991,diemand2008,vogelsberger2009,stadel2009,navarro2010}.  In fact, there are hints of tension with the neutralino paradigm on sub-galactic scales \citep{dobler2006,gentile2007,deblok2008,kuzio2008,kuzio2010,zwaan2010,boylan-kolchin2011,papastergis2011}. 
And yet, most of the effort to characterize the evolution of the Universe (experimentally, observationally, and theoretically) has been in the context of this dust-like cold dark matter (CDM).  In the absence of evidence for \emph{any} dark-matter candidate, much less the neutralino, it is important to explore the structure and evolution of the Universe for dark-matter phenomenology beyond cold and collisionless.

One intriguing possibility is that the dark matter belongs to and interacts with a ``dark'' or ``hidden'' sector \citep{khlopov2006,feldman2007,foot2007,pospelov2008,feng2008,arkanihamed2009,feng2009,sigurdson2009,cohen2010}.  The Standard Model has rich intra-sector phenomenology; it is not unreasonable to speculate that a complex dark sector only tenuously connected to the Standard Model might exist as well.  The simplest kind of dark-matter interaction is a hard-sphere interaction of identical dark-matter particles.
Such an interaction -- with an isotropic, velocity-independent, elastic scattering cross section -- was first introduced in an astrophysical context by \citet{spergel2000}.  Interactions of this type were invoked to ameliorate the tensions between observations and CDM predictions on small scales (on the scales of individual dark-matter halos) while leaving the large-scale successes of CDM intact.
In this paper we revisit this basic class of self-interacting dark matter (SIDM) using cosmological simulations to explore its effect on dark matter halo shapes as a function of cross section.  In a companion paper (Rocha et al. 2012) we investigate implications for dark matter halo substructure and density profiles.

We are reinvestigating this simple SIDM model, which had been decreed ``uninteresting'' in several studies a decade ago, for two primary reasons.  First, we suspected that the constraints 
that indicated that the SIDM cross section was too small to meaningfully alter the morphology of dark-matter halos, were not as tight as claimed.  Second, there is a wealth of new data (e.g., from near-field cosmology, lensing studies of galaxies and clusters) that may be better places to either look for SIDM or constrain its properties.  In this paper and our companion paper, Rocha et al. (2012), we reevaluate past constraints on SIDM and suggest several new places to look for the effects of SIDM on halo structure. 

There was a burst of work on SIDM before its untimely demise \citep{yoshida2000,yoshida:weak:2000,kochanek2000,hogan2000,dalcanton2001,dave2001,colin2002,hennawi2002}.  The death of isotropic, velocity-independent, elastically scattering SIDM largely came from the interpretation of three types of observations: halo evaporation in galaxy clusters \citep{gnedin2001}, cores in galaxy clusters \citep{yoshida:weak:2000,meneghetti2001}, and halo shapes \citep{miralda2002}.  The Y2K-era constraints from the former two classes of observations were at the level of $\sigma/m \lesssim 0.3\hbox{ cm}^2/\hbox{g}$ and $\sigma/m\lesssim 0.1\hbox{ cm}^2/\hbox{g}$ respectively.  However, we show in a companion paper, Rocha et al. (2012), that the evaporation and cluster-core constraints are likely overestimated.  

The most stringent constraints on dark matter models with large isotropic, elastic self-scattering cross sections emerged from the shapes of dark-matter halos, in particular from lens modeling of the galaxy cluster MS 2137-23 by \citet{miralda2002}.  This massive galaxy cluster has a number of radial and tangential arcs within $\sim 200$ kpc of the halo center \citep{mellier1993,miralda1995}.  \citet{miralda2002} argued that
 self interactions should make dark-matter halos round within the radius $r$ where the local per-particle scattering rate equals the Hubble rate $\Gamma(r) = \, H_0$, or equivalently, where each dark-matter particle experiences one interaction per Hubble time.   The scattering rate per particle as a function of $r$ in a halo scales in proportion to the local density and velocity dispersion, 
\begin{eqnarray}\label{eq:gamma}
 \Gamma(r) \sim \rho(r) (\sigma/m) v_{\mathrm{rms}}(r) \, ,
\end{eqnarray}
where $\rho$ is the local dark-matter mass density and $v_{\mathrm{rms}}$ is the rms speed of dark-matter particles.    Using the fact that the lens model needs to be elliptical at $70 \hbox{ kpc}$,  \citet{miralda2002} set a constraint of $\sigma/m \lesssim 0.02\hbox{ cm}^2/\hbox{g}$ on the velocity-independent elastic scattering cross section.  This constraint is one to two orders of magnitude tighter than other typical constraints on velocity-independent scattering \citep{yoshida:weak:2000,gnedin2001,randall2008}.  It rendered velocity-independent scattering far too small to form cores in low surface brightness galaxies and other small galaxies \citep{kuzio2008,deblok2008}. This is unfortunate because the main reason SIDM was interesting at the time was that it was a mechanism to create cores in such galaxies \citep{spergel2000}.

The tightness of the SIDM constraints on cluster scales has meant that the focus of SIDM studies has shifted to those on velocity-dependent cross sections, such that SIDM may significantly alter dwarf-scale or smaller dark-matter halos while leaving cluster-mass halos largely untouched \citep{feng2009,buckley2010b,loeb2011,vogelsberger2012}.  In recent times, such velocity-dependent interactions have arisen in hidden-sector models designed to interpret some charged-particle cosmic-ray observations  as evidence for dark-matter annihilation \citep{pospelov2008,fox2009,arkanihamed2009,feng2010d}.  Constraints on other hidden-sector dark-matter models have been made using X-ray isophotes of the gas in the halo of the elliptical galaxy NGC 720 \citep{buote2002,feng2009,feng2010d,buckley2010b,ibe2010,mcdermott2011,feng2012}.  

However, as we show below, reports of the death of isotropic, velocity-independent elastic SIDM are greatly exaggerated.  
In this paper, we show that these earlier studies did not correctly account for the fact that the observed ellipticity (of the mass in cylinders or the projected gravitational potential) gets contributions from mass well outside the core, and the region outside the core retains its triaxiality. We also show that for ellipticity estimators that are relevant observationally, there is significant amount of scatter and the overlap between CDM and SIDM ellipticities is substantial even for $\sigma/m=1\hbox{ cm}^2/\hbox{g}$. Lastly, we find that in the regions where SIDM particles have suffered (on average) about one or more interactions, the residual triaxiality is larger than what has been previously estimated \citep{dave2001}.  Along with the analysis in Rocha et al. (2012), we find that studies of the central densities of dark-matter halos are likely to yield tighter constraints on the SIDM cross section than the morphology of the halos.

We briefly summarize our simulations in Sec. \ref{sec:simulations}.  We present results on the three-dimensional shapes of  SIDM dark-matter halos compared to their CDM counterparts in Sec. \ref{sec:3dshape}.  We reexamine the previous SIDM constraints based on halo shapes in light of our simulations in Sec. \ref{sec:obs}.  In particular, we reexamine the \citet{miralda2002} constraint in Sec. \ref{sec:miralda} and from the shapes of the X-ray isophotes of NGC 720 \citep{buote2002} in Sec \ref{sec:xray}.  In Sec. \ref{sec:lensing}, we show how other lensing data sets may constrain SIDM in the future.  We summarize the key points of this paper and present a few final thoughts in conclusion in Sec. \ref{sec:conclusion}.

\begin{table}
\label{sims.tab}
\centering
\noindent {\bf Table 1:}  Summary of simulations.\\
\begin{tabular}{|l|cccc|}
\hline
Name & $L_\mathrm{Box}$&  $m_\mathrm{p}$ & $\epsilon$ & $\sigma/m$\\
&   [Mpc/h] &   [M$_\odot$/h] &
  [kpc/h] &  [$\hbox{cm}^2/\hbox{g}$]  \\
\hline \hline
CDM & $50$ &  $6.9\times10^7$ & $1.0$ &  -- \\
   & $25$ &  $8.6\times10^6$ & $0.4$ &  -- \\ 
   \hline
SIDM$_{0.1}$ & $50$  & $6.9\times10^7$ & $1.0$ & 0.1\\
                      & $25$  & $8.6\times 10^6$ & $0.4$  & 0.1\\
\hline
SIDM$_1$   & $50$ & $6.9\times10^7$ & $1.0$ &  1.0\\
                      & $25$ &  $8.6\times 10^6$ & $0.4$ &  1.0\\
\hline
\end{tabular}
\vskip 0.5 cm *Note: columns give name, simulation box size, particle mass, force resolution, and interaction cross section. We use $h=0.71$.  See Rocha et al. (2012) for more details on the simulations.  
\end{table}

\begin{figure*}
\begin{center}
\includegraphics[width=1.0\textwidth]{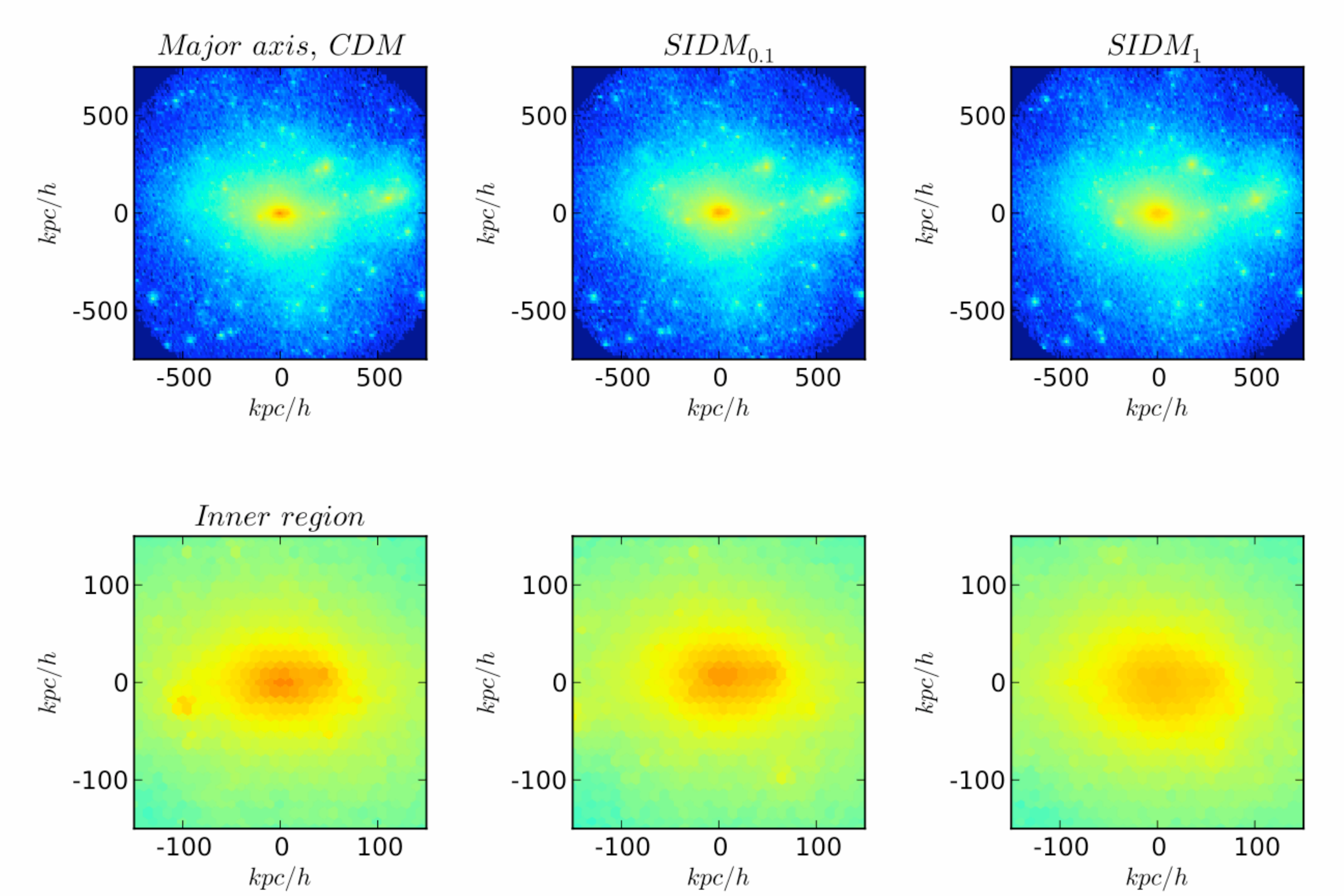}
\caption{\label{fig:major4}Surface density of a halo of mass $M_{\mathrm{vir}} = 1.2\times 10^{14}M_\odot$ projected along the major axis of the moment-of-inertia tensor -- the orientation that dominates the lensing probability.  The left column shows the halo for CDM, while the middle and right columns show the same halo simulated using SIDM with $\sigma/m = 0.1$ $\hbox{cm}^2/\hbox{g}$ and $1.0$ $\hbox{cm}^2/\hbox{g}$, respectively.  The bottom row shows the same information, now zoomed in on the central region. The surface density stretches logarithmically from $\approx 10^{-3}\hbox{g}/\hbox{cm}^2$ (blue) to $\approx 10\hbox{ g}/\hbox{cm}^2$ (red).}
\end{center}
\end{figure*}

\begin{figure*}
\begin{center}
\includegraphics[width=0.95\textwidth]{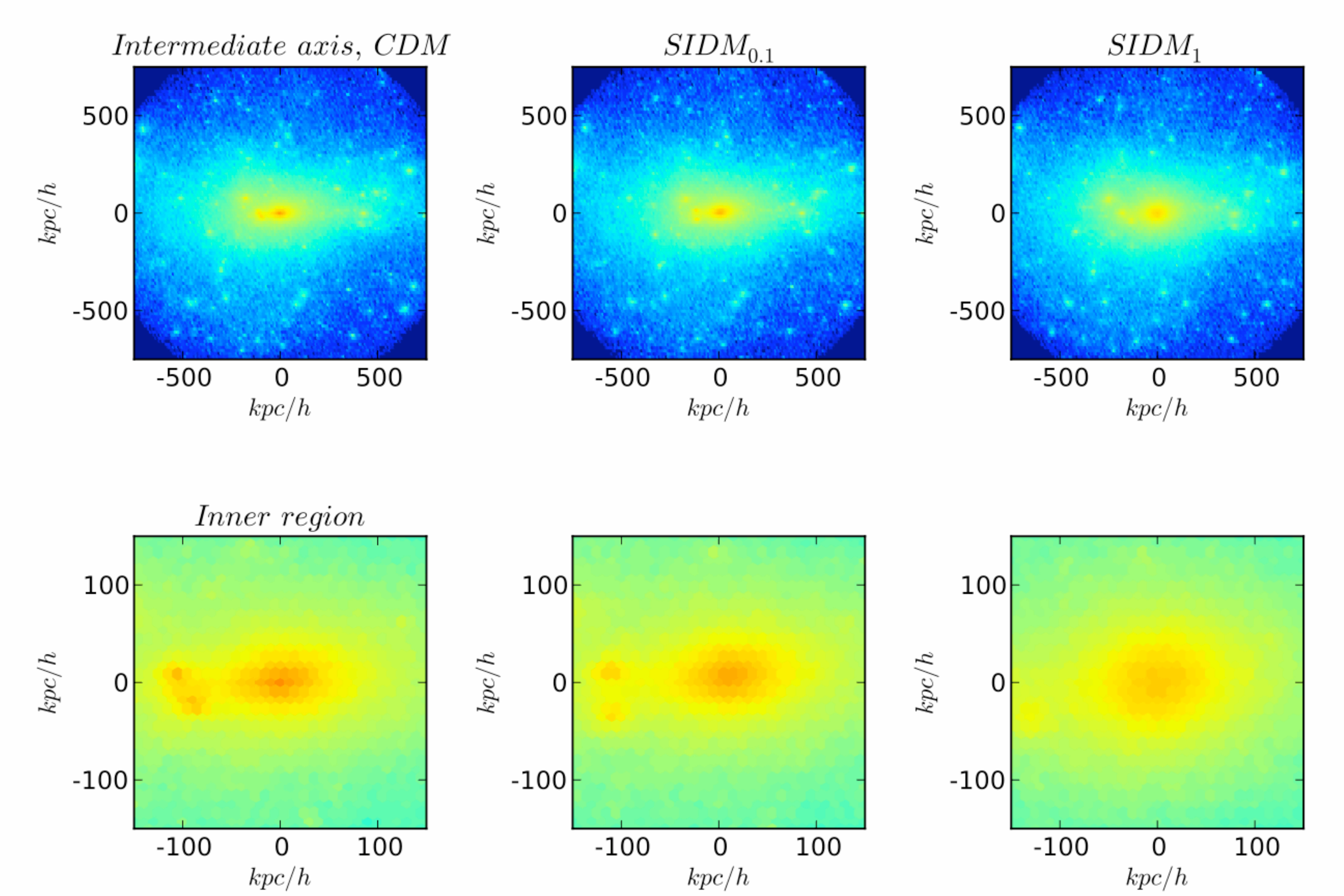}
\caption{\label{fig:intermed4} Surface density profiles for the same halo shown in Fig. \ref{fig:major4}, now projected along the intermediate axis.  Deviations from axisymmetry are highest along this projection.}
\end{center}
\end{figure*}

\section{Simulations}\label{sec:simulations}
We modeled self interactions by direct-simulation Monte Carlo with a scattering algorithm derived in Appendix A of Rocha et al. (2012) and implemented within the GADGET-2 \citep{springel2005} cosmological N-body code.  Once the code passed accuracy tests, we performed cosmological simulations of CDM and SIDM with identical initial conditions for cubic boxes of $25 h^{-1}$Mpc on a side and $50 h^{-1}$Mpc on a side, each with $512^3$ particles.  For the SIDM runs we
explored cross sections of $\sigma/m= 1,~0.1,~\hbox{and }0.03\hbox{ cm}^2/\hbox{g}$, though the lowest cross section run (with $\sigma/m = 0.03 \hbox{ cm}^2/\hbox{g}$) provided results that were so similar to CDM that we have not included them in any of the figures below.
The initial conditions  were generated using the MUSIC code at $z=250$ \citep{hahn2011} with a year seven WMAP cosmology \citep{komatsu2011}: $h = 0.71$, $\Omega_\mathrm{m} = 0.266$,
$\Omega_{\Lambda} = 0.734$, $\Omega_\mathrm{b} = 0.0449$, $n_\mathrm{s} =
0.963$, $\sigma_8 = 0.801$.  A summary of our simulation parameters, including particle mass and resolution is provided in Table 1.
We adopt a naming convention where the simulations denoted SIDM$_{0.1}$ and SIDM$_{1}$ have subscripts corresponding to their cross sections in units of $\hbox{cm}^2/\hbox{g}$.    In all cases the self-interaction smoothing length as defined in Rocha et al. (2012) was set to 2.8 times the force softening.

We locate and characterize halos using the publicly available Amiga Halo Finder \citep[AHF; ][]{knollmann2009} package.  The total mass of a host halo $M_{\mathrm{vir}}$ is determined as the mass within a radius $r_{\mathrm{vir}}$ using the virial overdensity as defined in \citet{bryan1998}.  Though most of our analysis focuses on distinct (field) halos, we also explore subhalo shapes.  For these objects their masses are measured within the radius at which the radial density profile of the subhalo begins to rise again because of the presence of the host. 

We present results on the shapes of halos at $z=0$ in a radial range $r_{\mathrm{min}}$ to $r_{\mathrm{vir}}$,  where $r_{\mathrm{min}}$ is the minimum radius within which we trust the shape measurements.  Since our two sets of simulations have different resolution, we can check the 
convergence of our shape estimates. 
For integral measures of shape (e.g., the moment-of-inertia tensor for all particles within a given radius), we find that although the density profiles look largely converged outside of the numerical-relaxation radius $r_{\mathrm{relax}}$ defined in \citet{power2003}, the shapes do not converge until at least $r_{\mathrm{min}} = 2 r_{\mathrm{relax}}$.  This radius is roughly $r_{\mathrm{min}} \approx 20\hbox{ kpc}$ for the halos in the $50h^{-1}\hbox{Mpc}$-sized simulations, with only modest dependence on halo mass and scattering cross section, and $r_{\mathrm{min}} \approx 10\hbox{ kpc}$ for the $25h^{-1}\hbox{Mpc}$ boxes.  Below $r_{\mathrm{min}}$ we find that the halo shapes are systematically too round.  However, shapes are more robust if found in shells (either spherical or ellipsoidal) because they are less contaminated by the effects of the overly round and numerically relaxed inner regions.  Shape estimates, especially integral estimates, are most reliable if there are at least $\sim 10^4$ particles within the virial radius (or tidal radius for subhalos), consistent to what has been found in earlier work \citep[e.g., ][]{allgood2006,vera-ciro2011}.

\begin{figure*}
\begin{center}
\includegraphics[width=0.95\textwidth]{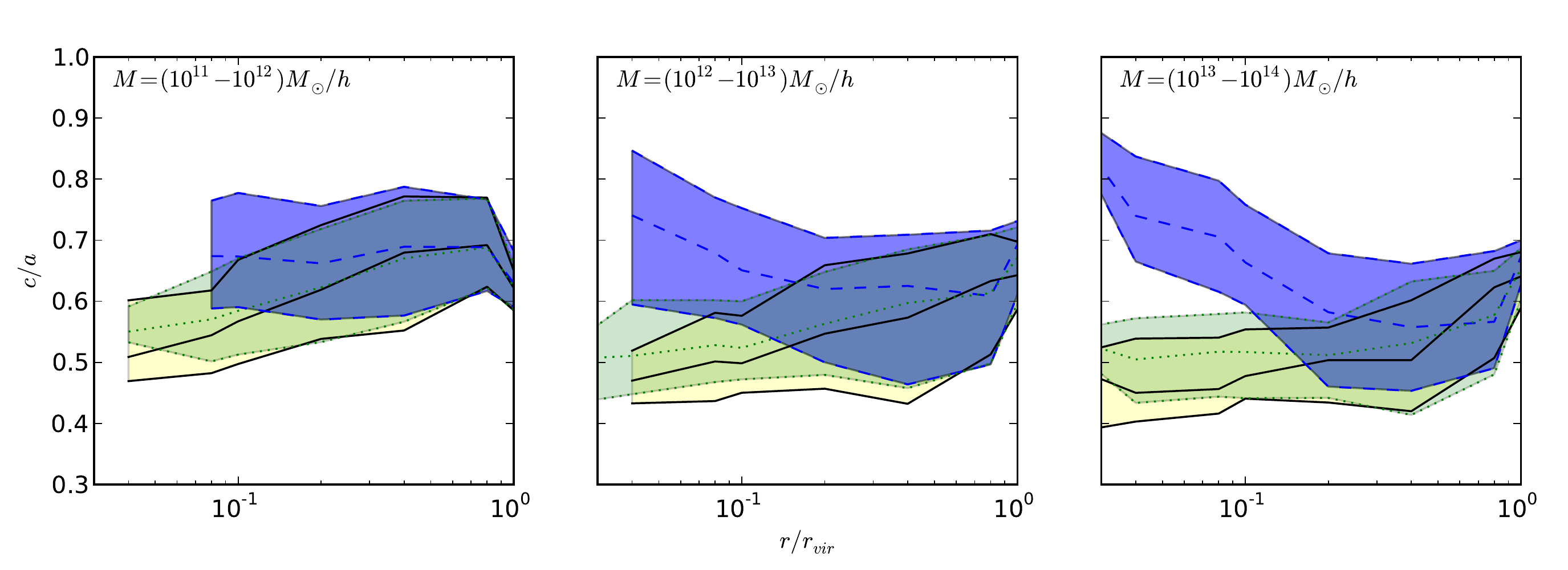}
\end{center}
\caption{\label{fig:3drrvir}Host halo shapes in shells of radius scaled by the virial radius in three virial-mass bins as indicated.    The black solid lines denote the 20th percentile (lowest), median (middle), and 80th percentile (highest) value of $c/a$ at fixed $r/r_{\mathrm{vir}}$ for CDM.    The blue dashed lines show the median and 20th/80th percentile ranges for $\sigma/m = 1\hbox{ cm}^2/\hbox{g}$, and the green dotted lines show the same for $\sigma/m = 0.1\hbox{ cm}^2/\hbox{g}$.  There are 440, 65, and 50 halos in each mass bin (lowest mass bin to highest).}
\end{figure*}

\section{Simulated halo shapes}\label{sec:3dshape}

\subsection{Preliminary Illustration}

Before presenting a statistical comparison of CDM and SIDM halo populations, we provide a pictorial illustration of  how an individual halo changes shape as we vary the cross section.  The columns of Figs. \ref{fig:major4} and \ref{fig:intermed4} 
show surface density maps for the same halo simulated in CDM, SIDM$_{0.1}$, and SIDM$_{1}$ from left to right.  In Fig. \ref{fig:major4}, we project the halo along the major axis, which is the orientation that maximizes the strong-lensing cross section \citep{vandeven2009,mandelbaum2009}.  In Fig. \ref{fig:intermed4}, we project the halo along the intermediate axis, which maximizes the deviation of the surface density from axisymmetry.  This particular halo is one of the most massive halos identified in the $50 h^{-1}$Mpc box runs, with $M_{\mathrm{vir}} = 1.2\times 10^{14}M_\odot$ and $r_{\mathrm{vir}} = 1.27\hbox{ Mpc}$.     The top row in each figure shows the surface density on the scale of the virial radius, while the lower row shows the inner $300h^{-1}$kpc of the halo (side-to-side).  The surface-density stretch is the same between the two figures.

The major and minor axes for the projections in these figures were determined using  the moment-of-inertia tensor of all particles within a sphere of radius $r_{\mathrm{vir}}$ in the halo.
If modeling the mass distribution as an ellipsoid, the principal axes $a (\mathrm{major}) > b (\mathrm{intermediate})> c (\mathrm{minor})$ are the square roots of the eigenvalues of this tensor.     

In comparing Figs. \ref{fig:major4} and \ref{fig:intermed4}, note that Fig. \ref{fig:major4} is the most relevant for strong-lensing studies (Sec. \ref{sec:obs}) and shows the smallest differences, especially at large radii. Indeed, only the zoomed view of the  $\sigma/m = 1\hbox{ cm}^2/\hbox{g}$ run is visibly rounder than the CDM case.  Even for the intermediate projection (Fig. \ref{fig:intermed4}), which maximizes the visual difference, the inner regions of the halo are only slightly rounder and less dense than their CDM counterparts for $\sigma/m = 0.1\hbox{ cm}^2/\hbox{g}$.  The $\sigma/m = 1\hbox{ cm}^2/\hbox{g}$ case is indeed less dense and \emph{rounder} within $\sim 100 h^{-1}\hbox{ kpc}$, but even in this case, some ellipticity is clearly evident.

A final point of interest in these visualizations concerns the substructure.   The subhalos apparent in the CDM halo are similarly abundant in the SIDM cases, and even approximately match in their positions.   There are minor differences in substructure densities and locations (especially in the central regions) but overall it is difficult to distinguish among the runs by comparing their substructure content (see Rocha et al. 2012 for a more quantitative comparison of substructure).

\subsection{Three-dimensional halo shapes}

We quantify halo shapes by examining ellipsoidal shells centered on the radial slices identified by AHF \citep{knollmann2009} for profile measurements.
We use shells instead of enclosed volume because it is less sensitive to numerical relaxation effects at the center and because it is a better estimate of the effects of local dark-matter scattering.  In each shell of material, we calculate a modified moment-of-inertia tensor \citep[defined and used in][]{allgood2006} in ellipsoidal shells, 
\begin{eqnarray}\label{eq:moi_ell}
  \tilde{\mathcal{I}}^{ell}_{ij}(a) & = & \sum_n^{\mathrm{shell}} \frac{x_{i,n}x_{j,n}}{r_n^2}, \quad \rm where\\
r_n & = & \sqrt{ x_{1,n}^2 + x_{2,n}^2/(b/a)^2 + x_{3,n}^2/(c/a)^2}\nonumber
\end{eqnarray} 
and $(x_{1,n}, x_{2,n}, x_{3,n})$ are the coordinates of the $n$th particle in the frame of the principal axes (major, intermediate, minor) of this tensor.  This is the same moment-of-inertia tensor from which shapes are inferred in \citet{dubinski1991} and \citet{dave2001}.
The principal axes ($a,b,c$) are computed as the square roots of the eigenvalues of $\tilde{\mathcal{I}}_{ij}^{ell}$.
The weighting of the moment-of-inertia tensor is chosen such that the outermost particles in the shell do not dominate the shape estimate.  We begin by finding the moment-of-inertia tensor in a spherical shell, setting $a=b=c=1$ for this initial estimate of $\tilde{\mathcal{I}}_{ij}^{ell}$, and iterate to find $\tilde{\mathcal{I}}_{ij}^{ell}$ with convergent $(a,b,c)$ values.  In each iteration, the ellipsoidal shell volume is defined using the ($a,b,c$) found in the previous iteration.  We experiment with either keeping the semi-major axis $a$ of the shell fixed between iterations or allowing $a$ to float such that the volume in the shell remains fixed as we iterate to find $\tilde{\mathcal{I}}_{ij}(a)$, but find that $c/a$ is insensitive to these choices.  Throughout this section, we show $c/a$ for fixed volume in the shell.  We only show results for $c/a$ because the trends for $b/a$ are similar but less informative, since $c/a$ indicates the deviation of the halo shape from sphericity.  

In order to understand trends, we split our analysis of host dark-matter halos and subhalos, and bin halos by virial (or tidal) mass.  Host halos are those whose centers do not lie within the virial radius of a more massive halo.  In Fig. \ref{fig:3drrvir}, we show the minor-to-major axis ratio $c/a$ as a function of radius normalized by the halo virial radius, $r/r_{\mathrm{vir}}$, for host halos in three mass bins.  Larger values of $c/a$ imply more spherical halos.    For the two lower mass bins, we used halos selected from the 25 $h^{-1}$Mpc boxes since these have the higher resolution.  For the highest mass bin, we used halos identified in the $50$ $h^{-1}$Mpc boxes in order to gain better statistics.   We checked to make sure that the results were convergent between boxes where the relevant mass resolutions overlap.  The shaded region corresponds to the 20th to 80th percentile for $c/a$ of the halo population for fixed $r/r_{\mathrm{vir}}$, and the central line shows the median value of $c/a$.  The black solid lines and yellow shaded regions denote shapes of CDM halos, the blue dashed lines and regions correspond to $\sigma/m = 1 \hbox{ cm}^2/\hbox{g}$, and the green dotted lines and regions correspond to $\sigma/m = 0.1\hbox{ cm}^2/\hbox{g}$.  The regions extend down in $r/r_{\mathrm{vir}}$ to the largest value of $r_{\mathrm{min}}/r_{\mathrm{vir}}$ in the given mass bin.

We reproduce the well-known trend that galaxy-mass halos in CDM are more spherical than cluster mass halos and that CDM halos become more spherical in their outer parts \citep{allgood2006}.  SIDM halos deviate most strongly from CDM at smaller radii, where the scattering rates are highest for a fixed cross section.     For SIDM$_{1}$,  halos are actually more spherical in their centers than their edges, with $c/a$ rising with decreasing $r$ for $r/r_{\mathrm{vir}} < 0.5$.  For SIDM$_{0.1}$, differences from CDM are only apparent for $r/r_{\mathrm{vir}} < 0.1$.  

\begin{figure}
\begin{center}
\includegraphics[width=0.45\textwidth]{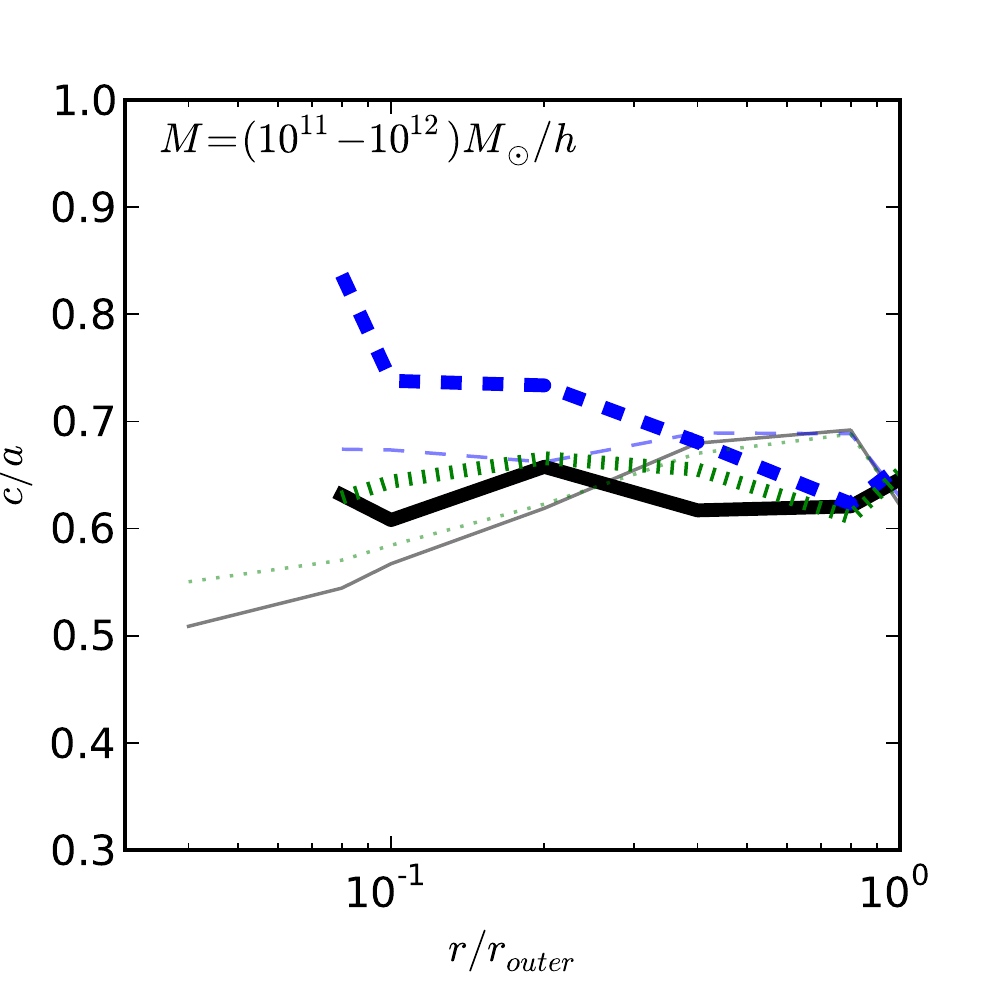}
\end{center}
\caption{\label{fig:3dsub}Median subhalo shape vs. radius for galaxy-mass systems compared to host halos of the same mass.  Bold lines denote subhalos and lighter lines denote host halos.  The radii are normalized by $r_{\rm vir}$ for hosts and $r_{\rm tidal}$ for subhalos.   Line colors and styles have the same meanings as in Fig. \ref{fig:3drrvir}:  dashed blue is SIDM$_{1}$, dotted green is SIDM$_{0.1}$, and solid black is CDM.}
\end{figure}

In Fig. \ref{fig:3dsub}, we compare the shapes of subhalos (thick lines) and host halos (thin lines) of similar mass by plotting the median axis ratio $c/a$ as a function of $r/r_{\mathrm{outer}}$ where $r_{\mathrm{outer}} = r_{\mathrm{tidal}}$ for subhalos (as defined by the AHF halo finder) and $r_{\mathrm{outer}} = r_{\mathrm{vir}}$ for host halos.  All halos have masses within $r_{\mathrm{outer}}$ between $10^{11} h^{-1}M_\odot$ and $10^{12} h^{-1}M_\odot$.  Though not shown, the 20th to 80th percentile ranges are similar in size as in Fig. \ref{fig:3drrvir}.    We find that the subhalo interiors in the $\sigma/m = 1\hbox{ cm}^2/\hbox{g}$ cosmology are systematically rounder than host halos.  
We speculate that there are are at least three effects that drive this trend.  
First, subhalos are typically more evolved than field halos of the same mass, with fewer recent mergers.  \citet{allgood2006} find that halos that form earlier are more spherical than halos that form later, which is attributed to directional merging, and more generally to the highly non-spherically-symmetrical way in which halos form and accrete.  Second,  
these subhalos are  the remains of more massive halos, which are more susceptible to the effects of self interactions at fixed $r/r_{\mathrm{vir}}$, as we showed in Fig. \ref{fig:3drrvir}.  Moreover, the outer radius of the subhalos is truncated with respect to its virial value, thus increasing $r/r_{\mathrm{outer}}$ for fixed $r$.  Since halos tend to be rounder in the outer parts with respect to the interiors for CDM halos, this boosts the initial $c/a$ for fixed $r/r_{\mathrm{vir}}$, beyond which SIDM boosts $c/a$ even more.  We see this trend for the CDM and $\sigma/m = 0.1\hbox{ cm}^2/\hbox{g}$ halos in Fig. \ref{fig:3dsub}.

\begin{figure*}
\begin{center}
\includegraphics[width=1.0\textwidth]{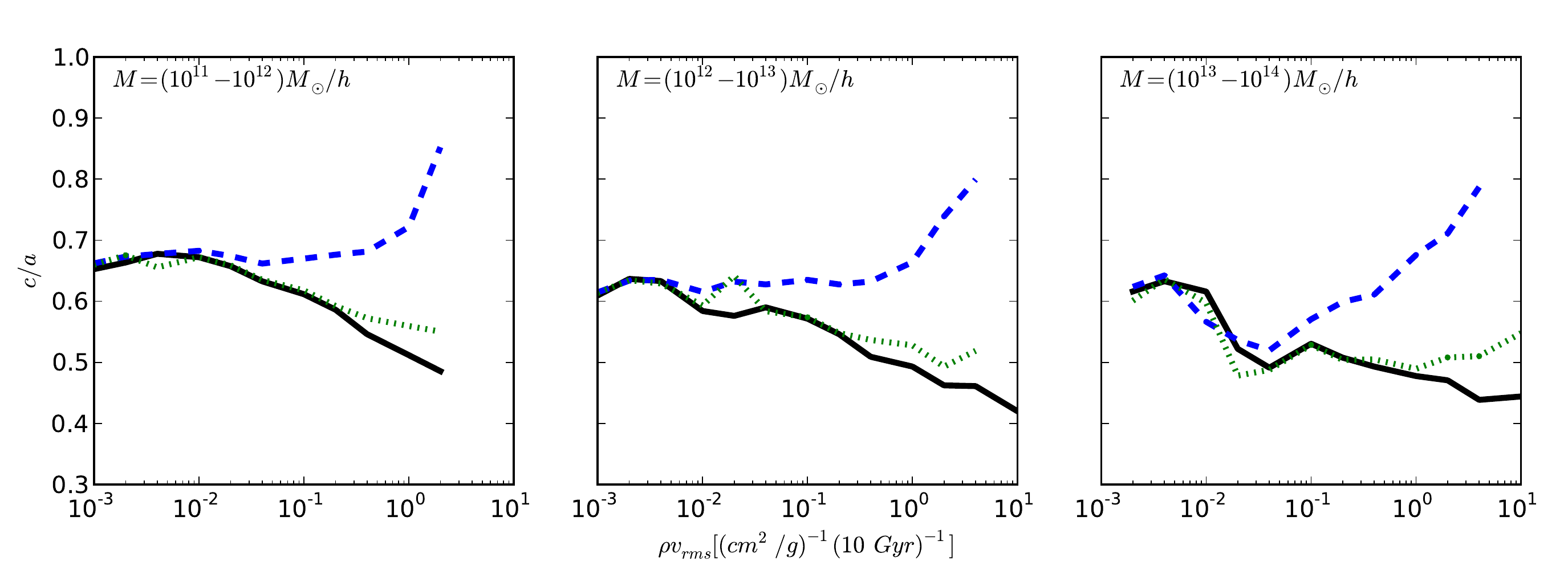}
\end{center}
\caption{\label{fig:3drhosig}Median axis ratio $c/a$ for host halos as a function of the local scattering rate modulo the cross section: $\rho v_{\mathrm{rms}} \sim \Gamma (\sigma/m)^{-1}$.  Smaller values correspond to the outer halo, where the density and scattering rate are low.  The quantity is scaled by $10 \hbox{ Gyr cm}^2/\hbox{g}$, such that $1$ in these units means that each particle has roughly one interaction per 10 Gyr in SIDM$_{1}$  (blue dashed line) and one interaction per 100 Gyr in SIDM$_{0.1}$ (green dotted line).  The black solid line is CDM.}
\end{figure*}

To see how the shape of dark-matter halos changes as a function of the typical local scattering rate (Eq. \ref{eq:gamma}), we plot $c/a$ as a function of  $\rho(r) v_{\mathrm{rms}}(r) \propto \Gamma(r) (\sigma/m)^{-1}$ in Fig. \ref{fig:3drhosig}.  The proportionality constant in relating  $\rho v_{\mathrm{rms}}$ to the scattering rate is ${\cal O}(1)$ and depends on the distribution function of dark matter particles. Thus it is reasonable to use $\rho v_{\mathrm{rms}}$ as a proxy for the local scattering rate modulo the actual cross section. To simplify the interpretation further, we multiply this quantity by $10 \hbox{ Gyr cm}^2/\hbox{g}$ in Fig. \ref{fig:3drhosig}.  In these units, if $\rho(r)v_{\mathrm{rms}}(r) > 1$ for $\sigma/m = 1 \, \hbox{cm}^2/\hbox{g}$, most particles will have scattered after 10 Gyr.  For $\sigma/m = 0.1 \, \hbox{cm}^2/\hbox{g}$, this quantity needs to be $10$ times larger to achieve the same scattering rate. Generally, $\rho v_{\mathrm{rms}}$ increases as one goes in towards the halo center, so particles tend to scatter more frequently in the core than in the outer parts of the halo, where interactions are uncommon over a Hubble time.

Fig. \ref{fig:3drhosig} shows that deviations in the halo shape from CDM begin when $\rho v_{\rm rms} (\sigma/m) \times (10\hbox{ Gyr})\sim 0.1$, independent of halo mass.  This corresponds to approximately 10\% of particles having scattered over a Hubble time at this radius.  However, the changes are small compared to the change where $\Gamma \times (10\hbox{ Gyr})\gtrsim 1$.  We note here that the most massive halos in \citet{dave2001} also seem to show the same qualitative behavior. 
However, even for large values of $\rho v_{\mathrm{rms}}$, the deviation from sphericity is significant, a fact that is in some disagreement with the simulation results of \citet{dave2001} where $c/a \gtrsim 0.9$ for their most massive halo. We speculate that part of this could be due to the differences in the way the ellipticity was estimated and part could be due to the smaller box run by \citet{dave2001}, which implies a quieter merger history.  
It is well known that CDM halos have anisotropic velocity ellipsoids and elongated shapes that are driven partially by directional merging \citep{allgood2006}.  These mergers provide a source of anisotropy that needs to be overcome by scattering in order for halos to reach sphericity.  We also note that the energy transfer facilitated by self-interactions would lead to an isotropic velocity dispersion tensor and that does not necessarily imply a rounder halo. To make this connection between isotropic velocity dispersion tensor and a rounder halo, previous analytic estimates have relied on the simulations of \citet{dave2001}.

The difference may also be a numerical artifact: we find that halo shapes only converge if there are at least $10^4$ particles in the halo and only for radii $r > 2r_{\rm relax}$ (see Sec. \ref{sec:simulations}). Most of the halos used for shape estimates in \citet{dave2001} only have $10^3$ particles in the virial radius.  Their largest halo does have more than enough particles for reliable shape estimates.  However, for this massive halo, \citet{dave2001} show shape measurements at radii much smaller compared to the convergence radius than we do.  In our simulations, we find that halos appear artificially round below the convergence radius. Our shape measurements are consistent with \citet{dave2001}'s massive halo for radii above the convergence radius.

\begin{figure}
\begin{center}
\includegraphics[width=0.45\textwidth]{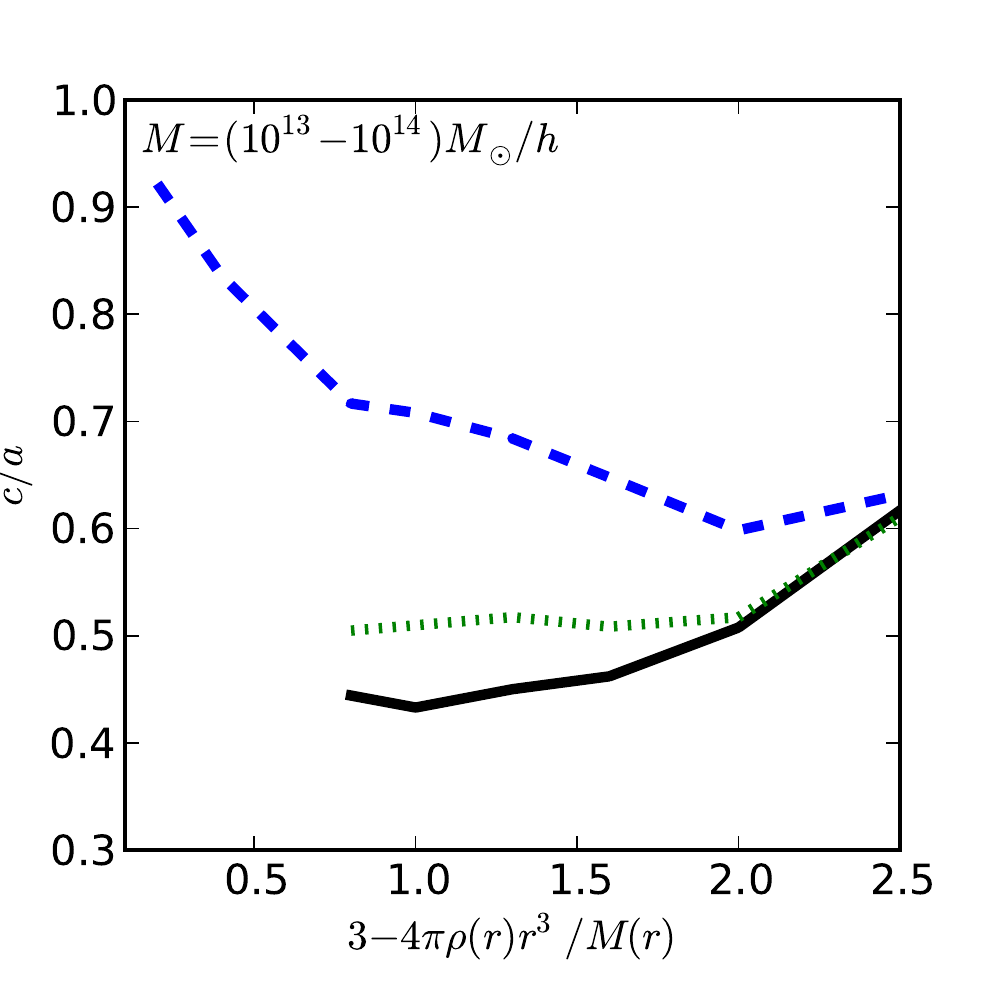}
\end{center}
\caption{\label{fig:3dlogslope}Median halo shapes as a function of the estimated log slope of the radial density profile.  Lines have the same meaning as in Fig. \ref{fig:3drrvir}.}
\end{figure}

The other major effect of energy transfers due to scattering is to create a core (see Rocha et al. (2012)) and deep inside a constant density core, we must have $c/a \rightarrow 1$. Hence we expect the log slope of the density profile to correlate strongly with the halo shape. 
Thus, instead of the density profile scaling as $\rho \sim r^{-1}$ in the interior as expected for CDM \citep{navarro1997,navarro2004,navarro2010}, the density profile of SIDM halos plateaus close to the center. We use the following proxy for the negative log slope of the density profile: 
\begin{eqnarray}\label{eq:logslope}
\gamma = 3 - 4\pi \rho(r) r^3/M(r), 
\end{eqnarray}
where $M(r)$ is the mass enclosed within radius $r$ and $\gamma=-d\log\rho/d\log r$ for a power-law density profile.  
Based on the previous figures, we expect halo shapes to become increasingly round as $\gamma \lesssim 1$, and that the CDM and $\sigma/m = 0.1\hbox{ cm}^2/\hbox{g}$ halos should not dip below $\gamma \approx 1$.  In Fig. \ref{fig:3dlogslope}, we show $c/a$ as a function of the log slope of the density profile, as approximated using Eq. (\ref{eq:logslope}), for host halos in our highest mass bin.  We obtain similar results for halos of smaller mass, but only show the highest mass bin because these halos are the best resolved.  

We find that, indeed, SIDM$_1$ halos become significantly rounder as $\gamma < 1$ and become almost completely round when $\gamma$ gets much smaller than 0.5. Interestingly, we also see that $c/a$ deviates strongly from CDM even for relatively large values for the log slope, in regions of the halo in which the scattering is not efficient at changing the radial density profile.  This is a consequence of the fact, as shown in Fig. \ref{fig:3drhosig}, that it does not take a lot of scatters to start rounding out the halos, although it takes multiple scatters for the halos to acquire $c/a$ axis ratios in excess 0.8.  Thus, the effects of scattering are apparent for $\sigma/m = 1\hbox{ cm}^2/\hbox{g}$ even when $\rho \sim r^{-2}$ and the density profile is unaffected by scatterings. However, the observational importance of this behavior is mitigated by two factors -- the change for $\gamma \gtrsim 1$ is mild and easily within the scatter in ellipticities seen in CDM. 

In summary, we find that it only takes a modest local scattering rate per particle, $\Gamma(r) \gtrsim 0.1 H_0$, to start changing the three-dimensional halo shape within radius $r$ with respect to CDM. We find that  SIDM$_1$ halos gets significantly rounder compared to CDM predictions when the negative log-slope of the density profile $\gamma \ll 1$ in regions where dark matter particles (on average) have had at least interactions in a Hubble time ($\Gamma \times (10\hbox{ Gyr})> 1$) . Our results show that even in the limit of one or more scatterings, the halo shapes retain some of their initial triaxiality. 

\section{Comparisons to Observations}\label{sec:obs}

\subsection{Defining Observables}

There are a number of ways of quantifying deviations of mass distributions from spherical or axial symmetry.  In the previous section, we quantified the deviations in terms of $c/a$, the ratio of the semi-minor to semi-major axes determined from the modified three-dimensional moment-of-inertia tensor, Eq. (\ref{eq:moi_ell}).  This is rarely a practical shape estimate observationally.  Instead, there is a more suitable measure of halo ellipticity or triaxiality for each type of observation.  The relationship between these measures is non-trivial, so care must be taken to compare theory to observation appropriately.  From paper to paper, the definition of  ``ellipticity" can change significantly.  In order to facilitate these comparisons, we define three distinct measures of asymmetry in this section and go on to use them for specific comparisons to observational studies in Sec. \ref{sec:miralda}, \ref{sec:xray}, and \ref{sec:lensing}.  The symbols we use for these shape definitions are summarized in Table 2.

\begin{table*}
\label{edef.tab}
\centering
\noindent {\bf Table 2:}  Summary of shape definitions used in observational comparisons\\
\begin{tabular}{llll}
\hline
Symbol & Defining Equation & Description & Relevant Figure \\
\hline \hline
$  e^\prime(R,\phi)$  & Eq. \ref{eq:eprime} & Deviation from axial symmetry in lensing convergence maps & Fig. \ref{fig:miralda} \\
\hline
$e$ & Eq. \ref{eq:e} & Ellipticity in dPIE surface density fits to lensing signal & Fig. \ref{fig:locuss} \\
\hline
$\epsilon$ & Eq. \ref{eq:epsilon} & Ellipticity of similar spheroid used in X-ray studies & Fig. \ref{fig:buote} \\
\hline
\end{tabular}
\end{table*}

For strong lensing, which we discuss in Secs. \ref{sec:miralda} and \ref{sec:lensing}, what matters is the deviation of the convergence from axial symmetry.
The convergence is $\kappa = \Sigma/\Sigma_{\mathrm{cr}}$, where $\Sigma_{\mathrm{cr}}$ is the critical surface density for creating multiple images.
For an axially symmetric system the convergence will be $\kappa(\theta)$, and depend only on the angular distance on the sky $\theta$ from then lens center.
Once axial symmetry is broken,  one must consider $\kappa(\theta,\phi)$, where $\phi$ is the azimuthal angle that rotates on the sky.  \citet{miralda2002} used the following quadrupole approximation to fit the surface density of MS 2137-23,
\begin{eqnarray}\label{eq:kappa}
  \kappa(\theta,\phi) = \kappa_0(\theta) - \frac{\varepsilon}{2} \frac{d\kappa_0}{d\theta} \cos(2\phi),
\end{eqnarray}
where $\kappa_0(\theta)$ is the convergence averaged along azimuthal angle and $\varepsilon$ quantifies the amplitude of deviation of the convergence from axial symmetry.  Since the normalization of $\kappa$ and the angle $\theta$ depend on the source, lens, and source-to-lens distances, which generically vary, we quantify deviations from axial symmetry in terms of surface density $\Sigma(R,\phi)$, where $R$ is the two-dimensional physical radius in projection.  Using Eq. (\ref{eq:kappa}), we define the measure of ellipticity
\begin{eqnarray}\label{eq:eprime}
  e^\prime(R,\phi) = \frac{2\left( \Sigma_0(R) - \Sigma(R,\phi) \right)}{d\Sigma_0/dR},
\end{eqnarray}
which should be equivalent to $\varepsilon \cos(2\phi)$ if the quadrupole expansion of the two-dimensional surface density is approximately correct.  Here, $\Sigma_0(R)$ is the azimuthally averaged surface density.

The second type of measure used to quantify deviations from spherical or axial symmetry in lensing arises from the extension of the double pseudo-isothermal sphere,
\begin{eqnarray}
  \rho(r) = \frac{\rho_0}{\left( 1 + r^2/r^2_{core}\right)^2\left( 1 + r^2/r_{cut}^2\right)^2},
\end{eqnarray}
to allow for deviations of the surface-mass density from axial symmetry \citep[double pseudo-isothermal elliptical, or dPIE; see ][]{richard2010},
\begin{eqnarray}\label{eq:dPIE}
  \Sigma_{dPIE}(x,y) =  
    \frac{\sigma_0^2}{2G} \frac{r_{cut}}{r_{cut} - r_{core}} \left[ \frac{1}{\sqrt{r_{core}^2 + \tilde{r}^2}} - \frac{1}{\sqrt{r_{cut}^2 + \tilde{r}^2}} \right],
\end{eqnarray}
with
\begin{eqnarray}\label{eq:e}
  \tilde{r}^2 = [(2-e)(x-x_c)/2]^2 + [(2-e)(y-y_c)/(2-2e)]^2.
\end{eqnarray}
The dPIE profile has the properties that $\rho \sim \hbox{const}$ for $r \ll r_{{core}}$, $\rho \propto r^{-2}$ for $r_{core} \ll r \ll r_{cut}$, and $\rho \propto r^{-4}$ for $r \gg r_{cut}$, so that the total mass is finite.  The center of the halo in projection is denoted by $(x_c, y_c)$, with the $x$-direction aligned with the major axis of the distribution.  This surface-density profile is often used in fits of the shapes of galaxies or dark-matter halos in clusters that strongly lens background galaxies.  

The third way we will quantify halo shapes observationally is in terms of similar spheroids \citep[see, e.g., Section 2.5 of ][]{binney2008}.  For spheroids, we can define an ellipticity parameter
\begin{eqnarray}\label{eq:epsilon}
  \epsilon = 1 - \frac{b}{a},
\end{eqnarray}
where $b$ is the semi-minor and $a$ the semi-major axis.  If the $z$-axis is the symmetry axis, the semi-major axis is given by 

\begin{eqnarray}\label{eq:obl}
  a^2 = R^2 + \frac{z^2}{(1-\epsilon)^2}, \hbox{ oblate}
\end{eqnarray}
for oblate spheroids, and
\begin{eqnarray}\label{eq:pro}
  a^2 = \frac{R^2}{(1-\epsilon)^2} + z^2, \hbox{ prolate}
\end{eqnarray}
for prolate spheroids. Similar spheroids are those for which the density profile may be described in terms of $\rho(a)$ and $\epsilon$ is fixed throughout the body.  The isopotential surfaces of such spheroids are rounder at large distances if the body is more centrally concentrated \citep[see the discussion in Sec. 2.5 of ][]{binney2008}.

As summarized in Table 2, $e^\prime$ is the surface-density shape definition we use in Sec. \ref{sec:miralda}, $\epsilon$ is the X-ray motivated shape definition we use in Sec. \ref{sec:xray}, and $e$ is the lensing-fit shape definition used in Sec. \ref{sec:lensing}.

\begin{figure*}
\begin{center}
\includegraphics[width=0.95\textwidth]{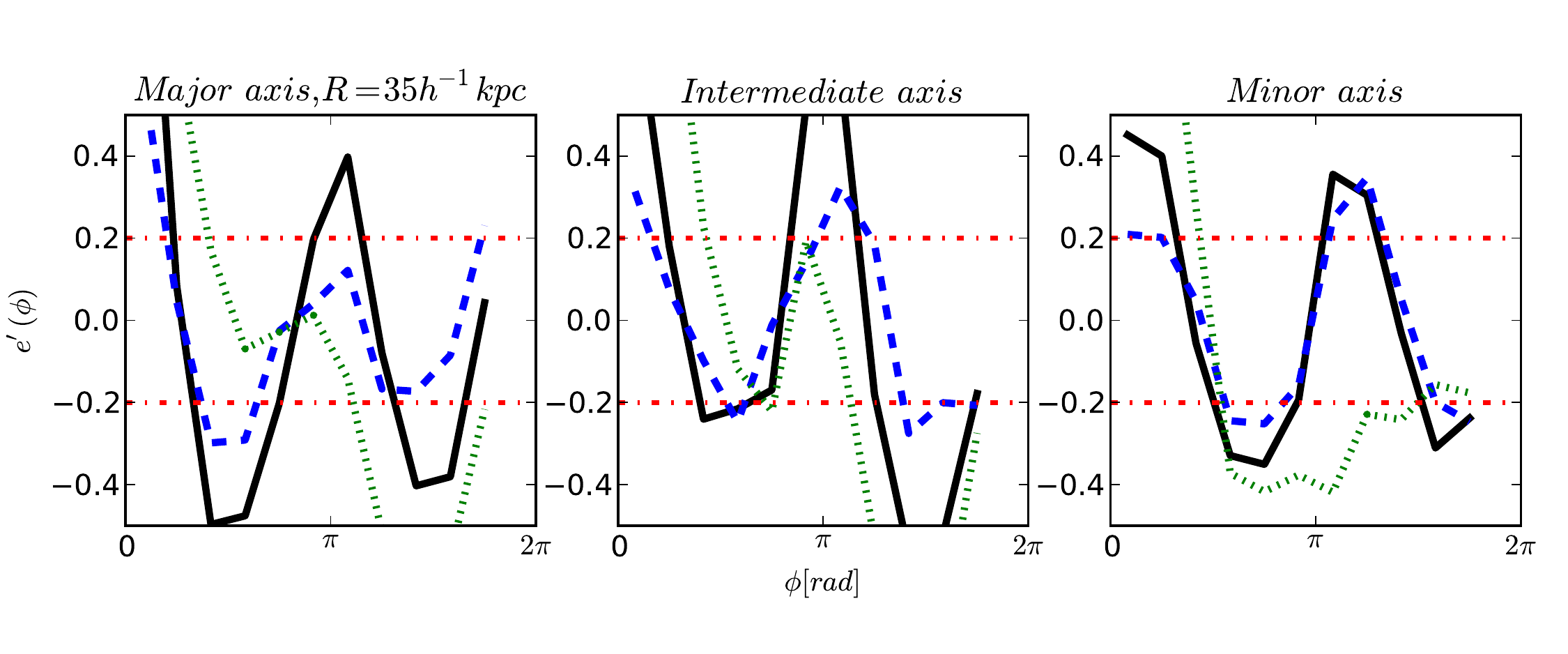}
\caption{\label{fig:miralda}The surface-density ellipticity $e^\prime(R,\phi)$ as defined in Eq. \ref{eq:eprime} for 
 the same halo as in Figs. \ref{fig:major4} and \ref{fig:intermed4}.  Shown are projections along the three principle axes, evaluated at $R=35$ kpc and plotted
 as a function of azimuthal angle.    As in the figures in Sec. \ref{sec:3dshape}, the solid black line corresponds to CDM, the blue dashed line to $\sigma/m = 1\hbox{ cm}^2/\hbox{g}$, and the green dotted line to $\sigma/m = 0.1 \hbox{ cm}^2/\hbox{g}$.  The dot-dashed red line shows the minimum amplitude of $e^\prime(\phi)$ required in order to explain the image positions of MS 2137-23.  We do not show the lines for the $\sigma/m = 0.03\hbox{ cm}^2/\hbox{g}$ simulation because it is indistinguishable from the CDM line.  The angle $\phi$ is defined such that $\phi = 0$ corresponds to the longer axis of the surface-density distribution.}
\end{center}
\end{figure*}

\subsection{Revisiting \citet{miralda2002}}\label{sec:miralda}
Galaxy clusters are great places to look for the effects of velocity-independent SIDM because one may typically achieve much higher values of $\rho(r)v_{\mathrm{rms}}(r)$ for fixed $r/r_{\mathrm{vir}}$.
In addition, there are many different probes of the mass distribution of clusters that span an enormous dynamic range of radial scale---stellar kinematics and strong lensing towards the center of the cluster, weak lensing and X-ray gas distributions throughout the halo volume, and weak + strong lensing maps of the matter distribution around individual galaxies in the cluster \citep{sand2008,newman2009,newman2011,kneib2011}.  It is no surprise the tightest constraints on velocity-independent SIDM emerged from cluster studies, and a revisit of the tightest of these constraints is the subject of this section.

The strongest published constraint on velocity-independent SIDM, $\sigma/m \lesssim 0.02\hbox{ cm}^2/\hbox{g}$ came from \citet{miralda2002}'s study of the galaxy cluster MS 2137-23.  This cluster has an estimated virial mass of $\sim 8 \times 10^{14}M_\odot$ \citep{gavazzi2005}, and its mass distribution has also been studied by \citet{fort1992}, \citet{mellier1993}, \citet{miralda1995}, \citet{gavazzi2003}, and \citet{sand2008}.  There are two strongly lensed galaxies that produce a total of five distinct images: one source has a radial image at $\theta \sim 5''$ from the center of the brightest cluster galaxy and an arclet at $\theta = 22.5''$.  The other source has a large tangential arc and two arclets all at about $\theta = 15''$ from the brightest cluster galaxy center, which corresponds to 70 kpc.  In order to reproduce both the relative magnifications and the alignments of the images in the sky, the surface density must deviate from axial symmetry at $70$ kpc.  Quantitatively, it means that the parameter $\varepsilon $ in Eq. (\ref{eq:kappa}), which corresponds to the amplitude of $e^\prime$ given in Eq. (\ref{eq:eprime}), must be $\varepsilon \approx 0.2$ at $R = 70$ kpc.  This figure is largely driven by the tangential arcs and associated arclets \citep{miralda1995}

Based on the ellipticity at $70$ kpc and an argument that the dark-matter surface density should be approximately axial for a typical particle collision rate $\Gamma \gtrsim H_0$, \citet{miralda2002} asserts that $\Gamma(70\hbox{ kpc}) \lesssim H_0$.  Using the fact that the tangential arc should lie at approximately where the mean interior convergence $\bar{\kappa} =1$ (or an estimated critical density $\Sigma_{cr} = 1 \hbox{ g/cm}^2$) and the rough approximation $\rho(r) \sim \Sigma(r)/r$, \citeauthor{miralda2002} estimates the three-dimensional density $\rho(70\hbox{ kpc})$.  Using the velocity dispersion of the brightest cluster galaxy at the center of the halo as a proxy for $v_{\mathrm{rms}}$, \citeauthor{miralda2002} uses Eq. (\ref{eq:gamma}) to determine a limit on $\sigma/m$, which is found to be $\sigma/m \lesssim 0.02\hbox{ cm}^2/\hbox{g}$.

We get a sense that this line of reasoning may be flawed when we examine the surface-density plots of one of our most massive halos in Figs. \ref{fig:major4} and \ref{fig:intermed4} and our findings of Sec. \ref{sec:3dshape}.  First, recall that our results show (cf., Fig.~\ref{fig:3drhosig})  that the inner halo shape retains some triaxiality even when $\Gamma \gtrsim H_0$. 
Second, the surface density includes all matter along the line of sight, not just the material within $r < R$.  
Thus, the surface density at small $R/r_{\mathrm{vir}}$ includes a lot of material with large $r$, far out in the halo where SIDM scatters are unimportant. This material is still quite triaxial.  Moreover, SIDM also creates cores, which means that the outskirts of the halo have an even greater weight in the total surface density than if the halo were still cuspy at the center.  Empirically, we see that the simulated surface densities in Figs. \ref{fig:major4} and \ref{fig:intermed4} are quite elliptical.  This point becomes more  and more important as the size of the core becomes smaller. All of these things suggest that the constraint reported in \citet{miralda2002} is far too high.

When attempting to quantify the constraints on SIDM from MS 2137-23 using our simulations, we run into the following problem: the largest halo in our simulations has a virial mass $M_{\mathrm{vir}} = 2.2\times 10^{14}M_\odot$, a factor of approximately four smaller than the estimated virial mass of MS 2137-23.  Moreover, we do not know the orientation of the principal axes of the cluster with respect to the line of sight.  In order to make the comparison, we do two things.  First, we can use virial scaling relations to estimate the radius at which $\rho(r) v_{\mathrm{rms}}(r)$, a proxy for the scattering rate (see Fig. \ref{fig:3drhosig}), has the same value as it would for a radius of 70 kpc in MS 2137-23, in other words, we look for the radius at which the SIDM scattering rate should be comparable to the radius at which \citet{miralda2002} finds the constraint for MS 2137-23.  For our $M_{\mathrm{vir}} \sim (1-2)\times 10^{14}M_\odot$ halos, $r = 35$ kpc is roughly the point at which the scattering rate is similar to that at 70 kpc in MS 2137-23.  This is outside $r_{\mathrm{min}}$ for these halos, so we trust the shape measurements.  Second, we look at several projections of the halos.  We calculate $\Sigma_0(R)$, $\Sigma(R,\phi)$, and hence $e^\prime(\phi)$ for the various projections of the halos.

We show an example of $e^\prime(\phi)$ curves for lines-of-sight along the principal axes of the halo moment-of-inertia tensors of one of our largest halos in Fig \ref{fig:miralda}, the same halo shown in Figs. \ref{fig:major4} and \ref{fig:intermed4}.  As in previous figures, the solid black line denotes the CDM result, the blue dashed line the result for $\sigma/m = 1\hbox{ cm}^2/\hbox{g}$, and the green dotted line the result for $\sigma/m = 0.1\hbox{ cm}^2/\hbox{g}$.  We do \emph{not} show the curve for $\sigma/m = 0.03\hbox{ cm}^2/\hbox{g}$ even though it is the cross section closest to the \citet{miralda2002} constraint because it is indistinguishable from the CDM line.  The red dotted line shows the minimum amplitude of $e^\prime(\phi)$ required for the lens model of MS 2137-23.  The $e^\prime(\phi)$ curves of the other massive halos look similar to the curves for the halo shown in Fig. \ref{fig:miralda}.  What we find is that even the $\sigma/m = 1\hbox{ cm}^2/\hbox{g}$ curve generally satisfies the MS 2137-23 constraint.  Therefore, we find that the \citet{miralda2002} constraint is in fact overly constraining by \emph{two orders of magnitude}.  While we do not simulate cosmologies with $\sigma/m > 1\hbox{ cm}^2/\hbox{g}$, and thus cannot set a quantitative upper limit on the SIDM cross section, we may conclude that $\sigma/m = 1\hbox{ cm}^2/\hbox{ g}$ is not ruled out by MS 2137-23.

There are a few caveats to this conclusion, which we do not believe will significantly alter our claim.  First, none of our simulated clusters are as massive as MS 2137-23. This precludes us from doing a detailed comparison of the projected densities in  $\sigma/m = 1\hbox{ cm}^2/\hbox{g}$ to that required to explain the arcs in MS 2137-23. It is, however, informative to do a simple calculation to gauge the importance of this effect. We appeal to the results in the companion paper (Rocha et. al. 2012) that show that the density profile in SIDM for $\sigma/m = 1\hbox{ cm}^2/\hbox{g}$ is well fit by a Burkert profile \citep{burkert1995}, and that this profile deviates from CDM density profiles at radii smaller than half the Burkert scale radius.  At this point, $r_\mathrm{b}/2$ or $0.35(r_\mathrm{max}/21.6\, \rm kpc)^{-0.08}$, the density profile becomes almost constant. We use the $V_\mathrm{max}-r_\mathrm{max}$ relation seen in our CDM simulations and the NFW profile to compute the projected mass within 70 kpc as the CDM prediction and compare that to the SIDM prediction by computing the same projected density but assuming that $\forall \, r<r_\mathrm{b}/2,\, \rho_{\rm SIDM}(r)=\rho_{\rm CDM}(r_\mathrm{b}/2)$. This computation reveals that the projected density in $\sigma/m = 1\hbox{ cm}^2/\hbox{g}$ model should be about 30\% lower than in CDM for the median $(1-5)\times 10^{15}\, \rm M_\odot$ halos. Even for the CDM case, however, the projected density is about a factor of 2 smaller than the estimated $\Sigma_{cr} = 1 \hbox{ g/cm}^2$. Clearly, the estimated projected density (as opposed to the shape) could be a significant constraint on the $\sigma/m = 1\hbox{ cm}^2/\hbox{g}$ model. This simple computation motivates further work along these lines with more realistic SIDM density profiles, inclusion of scatter and the uncertainty in the halo virial mass.

We have started much larger-scale simulations in order to study clusters in more detail, which will include an investigation of strong-lensing cross sections as well as ellipticity.  In terms of the ellipticity function $e^\prime(\phi)$, it is not clear in which direction our results will go for simulated $8\times 10^{14}M_\odot$ halos.  On one hand, large halos are more triaxial than small halos, so we would expect that if anything, we would be underestimating the degree of ellipticity in the surface-density distribution.  On the other hand, large halos have larger and rounder cores for velocity-independent SIDM compared to their lower-mass cousins.  This might drive the constraints the other way, although a larger core implies that the outskirts of the halo get weighted more heavily along the line-of-sight integral of the density for the surface-density calculation.  Finally, several authors have noted that MS 2137-23 is actually unusually round for a galaxy cluster \citep{gavazzi2005,sand2008}.  And intriguingly, the modeling of both \citet{miralda1995} and \citet{sand2008} indicate that a cored dark-matter radial density profile is preferred over a strongly, CDM-like cuspy profile for the dark-matter halo.  Such a cored profile is more in line with what we would expect for SIDM. 
Tighter constraints on $\sigma/m$, or a measurement should it be non-zero, will come from an ensemble of clusters, including the most triaxial of them, a point to which we return in Sec. \ref{sec:lensing}.

\begin{figure}
\begin{center}
\includegraphics[width=0.45\textwidth]{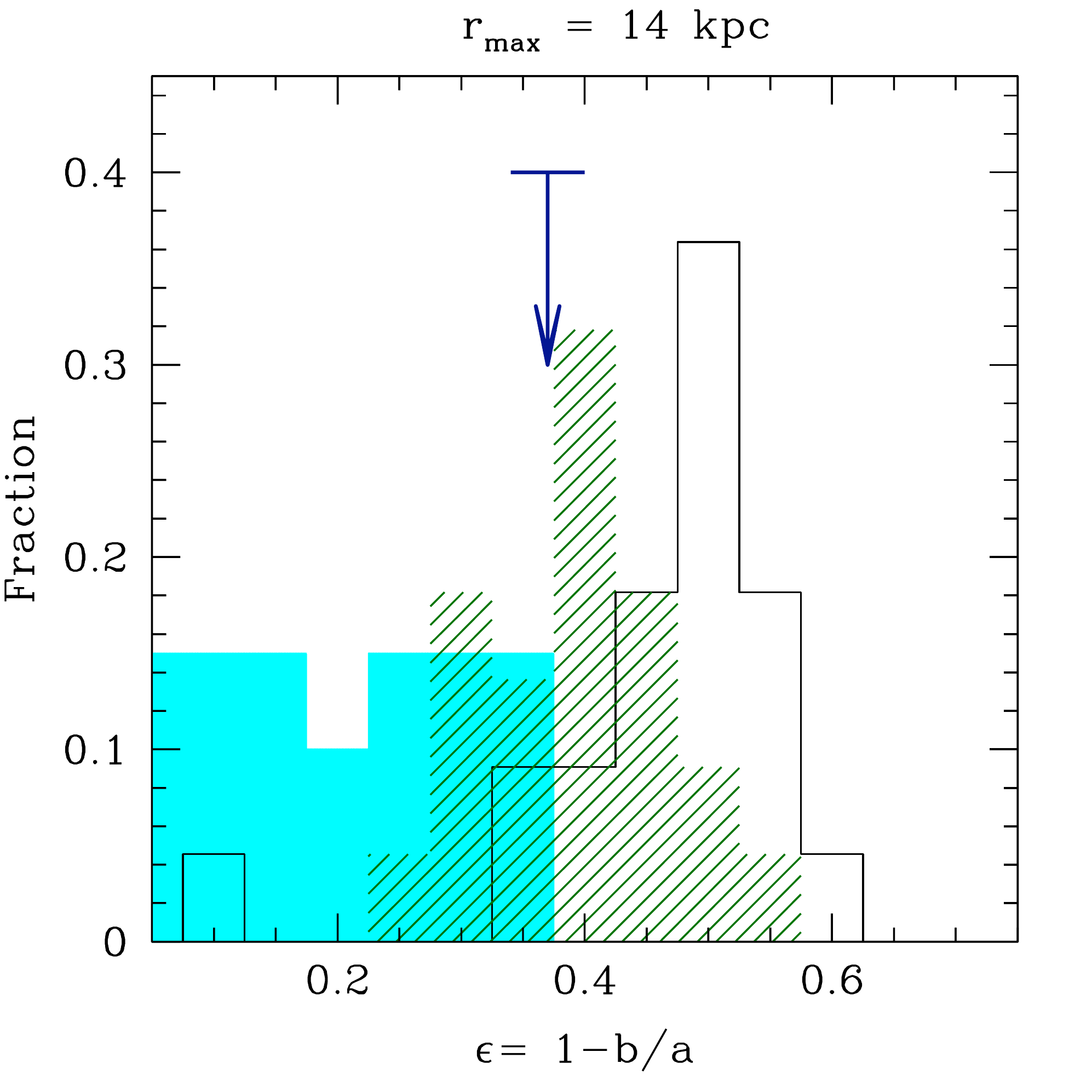}
\end{center}
\caption{\label{fig:buote} Distribution of halo ellipticity $\epsilon$ as defined in Eq. \ref{eq:epsilon} from fits for halos with  $M=(3-10)\times 10^{12} M_\odot$.  The black histograms indicate ellipticities for CDM halos, and the blue shaded histograms show the ellipticities for halos with $\sigma/m=1 \hbox{ cm}^2/\hbox{g}$, and green dashed histogram shows ellipticities if $\sigma/m = 0.1\hbox{ cm}^2/\hbox{g}$.  The dark blue line with arrow shows the center and width of the best-fit $\epsilon$ and uncertainty from \citet{buote2002}.  The shapes are found in using the \citet{dubinski1991} weighted moment-of-inertia method described in Sec. \ref{sec:simulations} and approximated as oblate or prolate spheroids.  The initial minimum radius of the shell for the shape measurement is set to $r_{\mathrm{min}} = 8.5$ kpc, and the maximum radius to $r_{\rm max} = 14 \hbox{ kpc}$.  See text for details.}
\end{figure}

\subsection{Shapes from X-ray observations of elliptical galaxies}\label{sec:xray}
Constraints on dark-matter scattering have also been made using the shapes of significantly smaller dark-matter halos, $M_{\mathrm{vir}} \sim 10^{12} - 10^{13} M_\odot$ \citep{feng2009,feng2010d,buckley2010b,mcdermott2011}.  In this case, the observations consist of X-ray studies of the hot gas halos of elliptical galaxies \citep{buote2002}.  

One may use the shape and twisting of the X-ray isophotes to learn about the three-dimensional shape of the dark-matter halo.  Interestingly, halo shapes may be constrained with imaging data alone and do not rely on temperature-profile modeling, which is required for mass-density determinations.  This ``geometric argument'' was first made by \citet{binney1978} and subsequently applied by a number of authors  in the study of elliptical galaxies and galaxy clusters \citep{fabricant1984,buote1994,buote1996,buote1998,buote2002}.  The geometric argument is the following: for a single-phase gas in hydrostatic equilibrium, the three-dimensional surfaces in the gas temperature $T$, gas pressure $p_g$, gas density $\rho_g$ and gravitational potential $\Phi$ all have the same shape \citep{binney1978,buote1998b}.  Surfaces of constant emissivity $j_X \propto \rho_g^2$ have the same three-dimensional shape as the isopotential surface.  

On the other hand, if the spectral data are available (as they typically are with the \emph{Chandra} telescope), then the temperature-profile data may be used with the assumption of hydrostatic equilibrium to fit the X-ray isophotes.  \citet{buote2002} use both approaches to model NGC 720.

X-ray isophotes are a good probe of the shape of the local matter distribution for the following reasons.  First, the total mass profile (galaxy + gas + dark-matter halo) is nearly isothermal ($\rho \sim r^{-2}$) for elliptical galaxies \citep{humphrey2006,gavazzi2007}.  This means that the shape of the isopotential contours reflects the shape of the matter distribution at the same radius, although typically the isopotential surfaces are significantly rounder than the isodensity contours \citep[see, e.g.,][]{binney2008}.  Second, the density profile of gas tends to be pretty sharply cuspy ($\rho_g \sim r^{-1.5}$).  Since the emissivity goes as $j_X \propto \rho_g^2$, and the morphology of $j_X$ traces that of the gravitational potential (see the discussion on the geometric argument above), this means that most of the X-ray emission along the line of sight is concentrated at radii close to the projected radius.  Thus, the X-ray isophotes indicate the shape of the matter distribution at radii similar to the projected radius.

In this section, we focus on the shape measurement of the X-ray emission about the elliptical galaxy NGC 720.  The inferred shape of the matter distribution of this system has been used to set constraints on self-interacting dark matter in the recent past \citep{feng2009,feng2010d,buckley2010b,mcdermott2011}.  The data set used by \citet{buote2002} for the shape measurement of NGC 720 is a 40 ks exposure of the inner $5^\prime$ of the galaxy with the ACIS-S3 camera on the \emph{Chandra} telescope.  The data included in the fit are contained within a 35$^{\prime\prime}-185^{\prime\prime}$ annulus from the center of the galaxy, which corresponds to $\approx 4.5-22.4$ kpc.  \citet{buote2002} estimate the three-dimensional isopotential shapes in the following way.  First, they investigate a mass-follows-optical light ($M\propto L_*$) mass profile for the gravitating mass using the geometric argument.  They use a spheroidal \citet{hernquist1990} model for the stellar mass of the galaxy, with structural parameters determined by deprojecting the optical image to three dimensions.  This is to test if the stars in the galaxy may be sufficient to provide a gravitational potential for the gas.  They find that the isophotes are rounder than observed beyond about $R_e$ ($\approx 50^{\prime\prime} \,\rm or\, 5 \rm\, kpc$).  This suggests that there must be significant ellipsoidally distributed mass extending well beyond the effective radius of stars, since isopotential contours become round as the distance to the main ellipsoidal mass profile increases.

Next, \citet{buote2002} use the fact that they find the temperature profile of the halo gas to be isothermal to find the three-dimensional X-ray emissivity distribution ($j_X \propto \rho_g^2$) directly from the equation of hydrostatic equilibrium:
\begin{eqnarray} \label{eq:jx}
   \frac{\rho_g(\mathbf{r})}{\rho_g(0)} &=& \exp \left[{-\frac{ \mu m_p \Phi(0)}{k_BT}\left( \Phi(\mathbf{r})/\Phi(0) - 1\right)} \right],
\end{eqnarray}
where $\mu$ is the molecular weight of the gas atoms and $m_p$ is the proton mass.  They used three similar spheroid models for the mass distribution of the galaxy (baryons plus dark matter) to find the gravitational potential $\Phi(\mathbf{r})$. The three models were pseudoisothermal ($\rho(a) \propto 1/ (1 + (a/a_p)^2)$ where $a$ is defined in Eqs. (\ref{eq:obl}-\ref{eq:pro})), Navarro-Frenk-White \citep{navarro1997} and Hernquist \citep{hernquist1990}.
They fit both oblate and prolate spheroids, and assumed that the symmetry axis was in the plane of the sky, as this leads to the most elliptical isophotes for a given spheroid.  Once the potential was found, they calculated the emissivity distribution in three dimensions and integrated along the line of sight.  A $\chi^2$ statistic was used to test the quality of the model fits to the X-ray surface-brightness map.

The best-fit models were the pseudoisothermal ones, with the oblate and prolate spheroids having nearly identical goodness-of-fit.  The best-fit ellipticity $\epsilon = 0.37 \pm 0.03$ for the oblate spheroid, and $\epsilon = 0.36 \pm 0.02$ for the prolate spheroid.  The flatness of the ellipticity as a function of radius in the region of interest drives the best-fit density-profile for isothermality.  Only for a similar isothermal spheroid are the X-ray isophotes so constant in ellipticity assuming density profiles that are similar spheroids.  

We compare our simulations to the \citet{buote2002} results by using the weighted moment-of-inertia tensor ellipsoidal shape estimator of Sec. \ref{sec:simulations}.  This method weights particles in the region of interest equally in estimating the shape of the mass distribution, and does not preferentially weight particles near the edge as the true moment-of-inertia tensor does.  The initial region in which the ellipsoidal shapes are estimated is a spherical shell of radius $r_{\mathrm{min}} < r < r_{\mathrm{max}}$, where $r_{\mathrm{min}}$  is our numerical limit on the shape convergence radius and $r_{\mathrm{max}} = 14 \hbox{ kpc}$.  We choose this value of $r_{\rm max}$ as a compromise between finding the ellipticity at small radii, where differences with CDM are highest, and having enough particles in the region of interest for a robust and unbiased shape estimate.  The shape of the region of interest is deformed in each iteration of the weighted moment-of-inertia calculation to reflect the major axes of the morphology of the tensor at that iteration, keeping the morphology of the region of interest ellipsoidal and fixing its volume.  This region is within the core radii of the $\sigma/m = 1\hbox{ cm}^2/\hbox{g}$ halos, and is our best approximation to the shape of the innermost part of the dark-matter halos, the parts relevant to the study of NGC 720.  For reference, the $10^{12}, 10^{13} \rm M_\odot$ halos have median core radii $16, 43$ kpc, respectively (Rocha et. al., 2012).

In order to find the \emph{spheroidal} ellipticity (described in Eq. (\ref{eq:epsilon})) from these ellipsoidal fits, we use the ellipsoidal axis ratios to decide if the halo is oblate or prolate.  If prolate, we take axis ratio of the spheroid $1-\epsilon$, cf. Eqs. (\ref{eq:obl}) \& (\ref{eq:pro}), to be $\sqrt{bc}/a$, i.e., the spheroidal semi-minor axis is the geometric mean of the ellipsoidal minor and intermediate axes.  If oblate, we set $1-\epsilon=c/\sqrt{a b}$, i.e., the semi-major axis of the oblate spheroid is set to be the geometric mean of the major and intermediate axes of the ellipsoid. Both these choices set the spheroidal volume equal to the ellipsoidal volume. A more realistic analysis would compute the integrated $j_X$ directly from the potential of the simulated halo and compare that to the observations. However, there are significant other uncertainties (discussed below) that argue against such an approach being more fruitful. 

In Fig. \ref{fig:buote}, we show the ellipticity distribution for all halos in our 25 $h^{-1}$Mpc boxes for CDM and SIDM$_1$ that have a virial mass within the 1$\sigma$ uncertainties of the mass modeling in \citet{humphrey2006} for NGC 720, $M_{\mathrm{vir}} = (6.6^{+2.4}_{-3.0})\times 10^{12}M_\odot$.  Note that we do not weigh the $\epsilon$ distribution by the error distribution for the virial mass presented in \citet{humphrey2006}.  However, the $\epsilon$ distribution in our simulated halos is not a strong function of virial mass in this mass range.  We show the CDM $\epsilon$ distributions with the black histograms, the $\sigma/m = 1\hbox{ cm}^2/\hbox{g}$ $\epsilon$ distributions with the cyan shaded histograms, and $\sigma/m = 0.1\hbox{ cm}^2/\hbox{g}$ with dashed green histograms.  We find that the inferred ellipticities are approximately independent of $r_{\rm max}$ out to approximately 25 kpc, with there being a small tail at higher $\epsilon$ for the $\sigma/m = 1\hbox{ cm}^2/\hbox{g}$ halos since the core radii are in the 20-40 kpc range and halo shapes are less affected by scatterings at radii larger than roughly the core radius.  For $\sigma/m = 0.1\hbox{ cm}^2/\hbox{g}$, the core sizes are of order $r_{\rm min}$ or smaller, so the shape measurements are relatively insensitive to $r_{\rm max}$ range.  A quick review of Fig. \ref{fig:3drrvir} also shows that the relative independence of the ellipticity estimate out to $0.1 r_{\rm vir}$ is to be expected in each cosmology.  For the $\sigma/m = 1\hbox{ cm}^2/\hbox{g}$ halos, we exclude those halos for which either poor centering of the halo or ongoing merging makes the halos appear artificially flattened.  For the other cosmologies, the relative cuspiness of the central density leads to more accurate halo centering.

Although there are significant differences in the ellipticity distributions of the CDM and SIDM halos, the observed ellipticity (downward arrow) is just within the ellipticity distribution for $\sigma/m = 1\hbox{ cm}^2/\hbox{g}$. 
The fact that this magnitude of ellipticity is still in the distribution for $\sigma/m = 1\hbox{ cm}^2/\hbox{g}$ is also apparent from the middle panel of Fig. \ref{fig:3drrvir}, which shows that the three-dimensional axis ratios at $r/r_{\rm vir} \simeq 0.04$ (approximately the radius corresponding to the 2D region of interest for NGC 720) can accommodate significant ellipticity. Note that radius is also close to the typical core radius where there we expect roughly 1 interaction per particle on average. 
This implies that NGC 720 could be an extreme outlier in SIDM$_1$ model and thereby consistent with observations. 

However, there are some issues with this interpretation. One issue is that the ellipticity measured for the dark matter halo seems to be constant down to roughly $\sim 5$ kpc in the data, which is about $1\%$ of the virial radius for the mean virial mass for NGC 720 \citep{humphrey2006}. The middle panel of Fig. \ref{fig:3drhosig} shows a sharp rise in axis ratio when the number of interactions gets above unity and hence our results are likely to change if we were able push deeper into the core with higher-resolution simulations. A second issue is that the measured halo mass for dark matter within 10 kpc \citep{humphrey2006} is a factor of 2-3 larger than even the median mass in our CDM simulations for halos with virial mass in the range preferred by \citet{humphrey2006} fits. Thus we should also cut the histogram in Fig. \ref{fig:buote} based on, say, the mass within 10 kpc (or some other measure of the concentration). These halos, by virtue of their larger densities, will also be rounder.  The discrepancy in average dark-matter density within 10 kpc may also be a sign that the inner part of the dark-matter halo has been compressed as a consequence of assembly of the elliptical galaxy, a compression that has been observed in other studies of elliptical galaxies \citep{schulz2010,dutton2011}.  Moreover the presence of the stars and gas increase the velocity dispersion of the dark matter in the central parts and increase the scattering rate -- an effect that is not captured in our SIDM simulations. 

 The resolution of these issues lies in a better comparison to the data, which in turn requires a bigger box higher-resolution simulation to probe deeper into the halo and gain more statistics. It will be important to include baryons to see how their presence may affect halo properties.  With such a halo catalog in hand, it will be interesting to do a more careful comparison to the X-ray data of NGC 720 and other large nearby ellipticals. The addition of other ellipticals would be crucial. With only one object, the spread we see in ellipticities may be hard to overcome (although it may be smaller if we also cut on concentration as discussed above). If we had an ensemble of shape measurements, we would be able to set tighter limits on the SIDM cross section. 

While our comparison to \citet{buote2002} is not sufficiently sharp, the weight of the arguments suggests that  $\sigma/m = 1\hbox{ cm}^2/\hbox{g}$ is not likely to be consistent with the measured shape of NGC 720.  However, based on our existing simulations, $\sigma/m = 0.1\hbox{ cm}^2/\hbox{g}$ is as consistent as CDM for the shape of the NGC 720 isophotes.  
It is interesting to note in this regard that there is no hint of a large core ($\sim 30$ kpc) in the results of \citet{buote2002} or \citet{humphrey2006}. The core sizes for $\sigma/m = 0.1\hbox{ cm}^2/\hbox{g}$ are smaller, $\sim 7$ kpc, for the same virial mass range (Rocha et al, 2012), comparable to the effective radius of the stars in NGC 720. 
Thus the inferred dark matter density profile in NGC 720 may be a better way to search for effects of self-interactions. 

To amplify the point about the dark matter density further, we note that the median central (maximum) density for SIDM with $\sigma/m = 0.1\hbox{ cm}^2/\hbox{g}$ is $0.05 \rm M_\odot/\rm pc^3$, while \citet{humphrey2006} infer an average density of $0.04 \rm M_\odot/\rm pc^3$ within 10 kpc. At 5 kpc, the inferred average density is $0.1 \rm M_\odot/\rm pc^3$, still within a factor of two (expected from scatter) of the predictions. This lends credence to the argument that an analysis focused on SIDM predictions of an ensemble of nearby X-ray-detected elliptical galaxies could be a fruitful way to look for signatures of or constrain SIDM. It is also worth noting that the shape distribution of $\sigma/m = 0.1\hbox{ cm}^2/\hbox{g}$ is visibly different from CDM and perhaps an ensemble of X-ray-shape measurements could resolve the differences. 

Our analysis argues for the conclusion that SIDM cross sections with $\sigma/m \gtrsim 0.1\hbox{ cm}^2/\hbox{g}$ can be hidden in X-ray data, and that previous constraints on SIDM using these data are overly stringent.  \citet{feng2010d} assumed that the \citet{buote2002} ellipticities described the halo shape at the inner radius of $R = 4.5\hbox{ kpc}$ and that $\Gamma \sim H_0$ was required to make the halo spherical as indicated by the results of \citet{dave2001}. Our results indicate that this interpretation is flawed because (a) SIDM halos retain significant triaxiality in the region where $\Gamma \sim H_0$ and (b) the scatter in SIDM halo ellipticities is large. In order to use analytic arguments to constrain self-interacting dark matter models, they should be tuned to reproduce the the distribution of axis ratios seen in simulations.

\subsection{The future of cluster lensing constraints}\label{sec:lensing}
In the future, far better constraints on SIDM will come from statistical studies of galaxy-cluster lens samples rather than the analysis of individual objects. In this section, we focus on statistical studies of the shapes of relaxed clusters. 

There are a number of ongoing and future observational programs designed to characterize the mass function of and mass distribution within galaxy clusters \citep[e.g., ][]{lsst2009,gill2009,plagge2010,richard2010,planck2011viii,planck2011ix,planck2011xii,pillepich2011,marriage2011,viana2012,postman2012}.  Modulo the effects of baryons, a smoking-gun sign of SIDM would be for the mass function of galaxy clusters to look identical to CDM, but with lower mass density and rounder surface-mass distributions at the centers of the clusters.  One would use the ensemble of galaxy-cluster data to compare with simulations of clusters in CDM and SIDM (with various elastic scattering cross sections). 
%

In this study, we will consider shape-based constraints from the initial results of the Local Cluster Substructure Survey (LoCuSS; PI: G. Smith), a multi-wavelength follow-up program of 165 low-redshift clusters selected from the ROSAT All-sky Survey catalog \citep{ebeling2000,boehringer2004,richard2010}.  Twenty clusters were used for the first mass-modeling study \citep{richard2010}.  These were selected because they had been observed with the Hubble Space Telescope (\emph{HST}), could be followed up spectroscopically at Keck, and were confirmed to have strongly lensed background galaxies.  The details of the mass modeling of these clusters are presented in Sec. 3 of \citet{richard2010}.  For our purposes, the key fact is that the surface density of the cluster-mass dark-matter-halo component of the lens model was parametrized in terms of the dPIE profile, Eq. (\ref{eq:dPIE}).  This study should be taken as an example of the power of using statistical studies of clusters for SIDM constraints, with an emphasis on the constraints on halo shapes.

In order to compare the LoCuSS clusters to simulations, we fit the surface densities of our most massive clusters in the 50 $h^{-1}$Mpc boxes for our
CDM, SIDM$_{0.1}$, and SIDM$_{1}$ runs using the dPIE surface-density profile, Eq. (\ref{eq:dPIE}).  We fit the surface densities of the halos projected along the principal axes of the moment-of-inertia tensors of the mass within the virial radius, and perform the fits within annuli $R_{\mathrm{min}}<R<R_{\mathrm{max}}$.  The inner radius of the annuli is chosen to be $R_{\mathrm{min}} = 20\hbox{ kpc}/h$, since this is the three-dimensional shape convergence radius (see Sec. \ref{sec:simulations}).  We set the outer radius $R_{\mathrm{max}} = 50\hbox{ kpc}/h$ since most of the lensing arcs in LoCuSS are in the range $20\hbox{ kpc} < R < 100\hbox{ kpc}$ if projected into the plane of the sky at the position of the galaxy clusters.  We fix $r_{cut} = 1000\hbox{ kpc}$, which is what is typically done with the LoCuSS clusters.  In practice, the parameter constraints are insensitive to this choice, as all the data are well within this radius if projected onto the sky.  The free parameters that we fit are: $\sigma_0$, the characteristic velocity dispersion of the cluster; $r_{core}$, the pseudoisothermal core radius; $e$, the cluster ellipticity (as defined in Eq. (\ref{eq:e})); and the position angle $\theta$.  We fix the center of the cluster to that inferred by AHF.  The surface-density parameters were fit using the downhill simplex algorithm in the \texttt{scipy} \texttt{python} module using the likelihood
\begin{eqnarray}
  \mathcal{L} = P(N_{obs}|N(\sigma_0,r_{core},e,\theta)) \prod_i^{N_{obs}} \mathcal{L}_i,
\end{eqnarray}
where $P(N_{obs}|N(\sigma_0,r_{core},e,\theta))$ is the Poisson probability of finding $N_{obs}$ simulation particles within the annulus given a dPIE model with parameters $(\sigma_0,r_{core},e,\theta)$, for which $N$ simulation particles are expected.  $\mathcal{L}_i$ is the probability of finding a simulation particle $i$ at position $(x,y)$ given those same parameters,
\begin{eqnarray}
  \mathcal{L}_i = \frac{\Sigma(x,y|\sigma_0,r_{core},e,\theta)}{\int_{\mathrm{annulus}} dxdy \Sigma(x^\prime,y^\prime)},
\end{eqnarray}
where $\Sigma(x,y)$ is defined in Eq. (\ref{eq:dPIE}).

\begin{figure*}
\begin{center}
\includegraphics[width=0.95\textwidth]{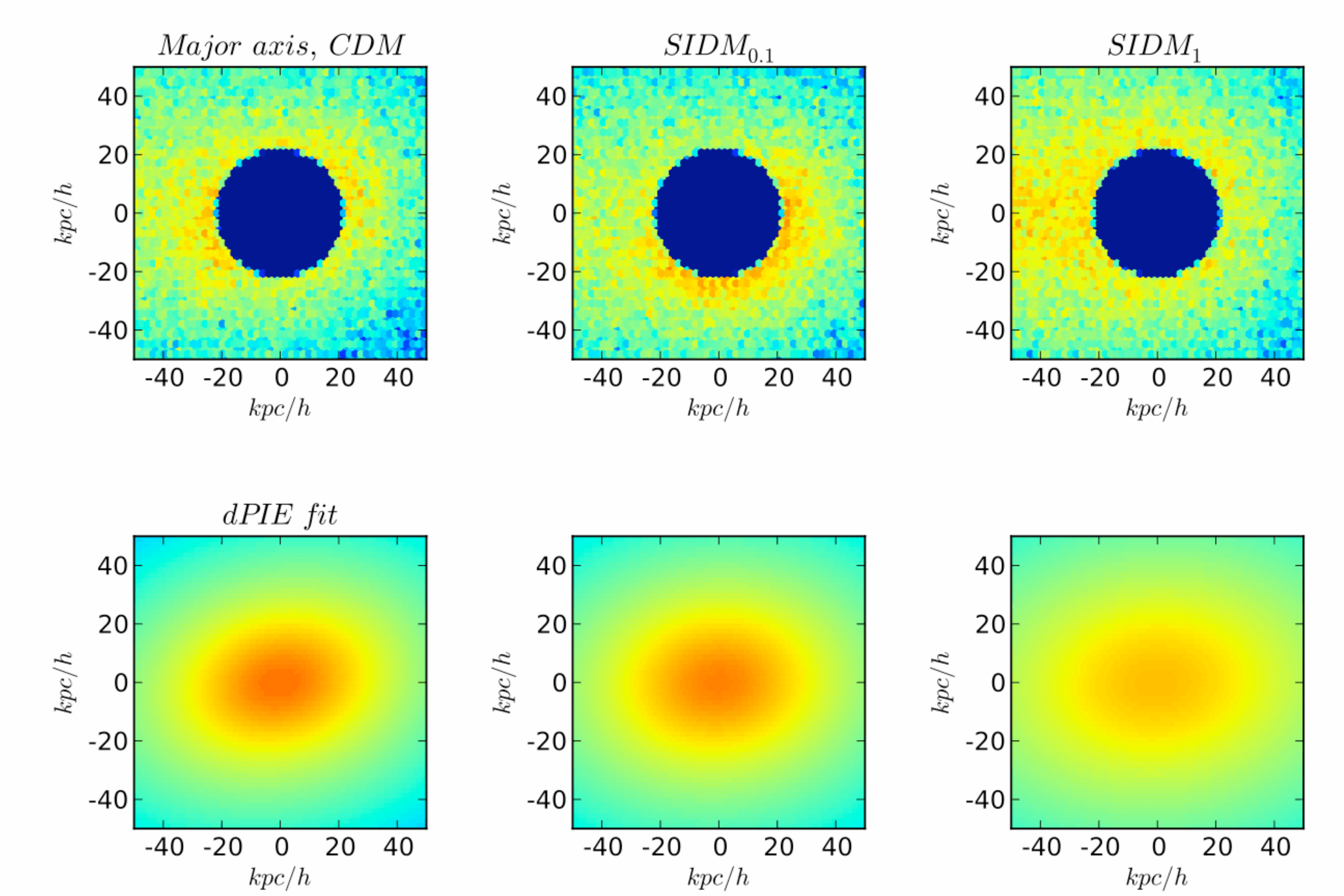}
\end{center}
\caption{\label{fig:dPIE_CDM} \emph{Top row:} Surface density of a simulated galaxy cluster of mass $M_{\mathrm{vir}} = 1.8\times 10^{14}M_\odot$ in CDM (left), $\sigma/m = 0.1\hbox{ cm}^2/\hbox{g}$ (center) and $\sigma/m = 1\hbox{ cm}^2/\hbox{g}$ (right) cosmologies. The central regions in which the projected radius $R < r_{\mathrm{min}}$ are masked as they are in the fits.  These surface densities result from viewing the halo along the major axis of the moment-of-inertia tensor of all particles in the halo.  \emph{Bottom row:} Best-fit dPIE surface-density fits for the surface densities above.  The ellipticities of the dPIE fits are: CDM, $e = 0.28$; $\sigma/m = 0.1\hbox{ cm}^2/\hbox{g}$, $e=0.25$; $\sigma/m = 1\hbox{ cm}^2/\hbox{g}$, $e=0.29$. }
\end{figure*}

\begin{figure*}
\begin{center}
\includegraphics[width=0.95\textwidth]{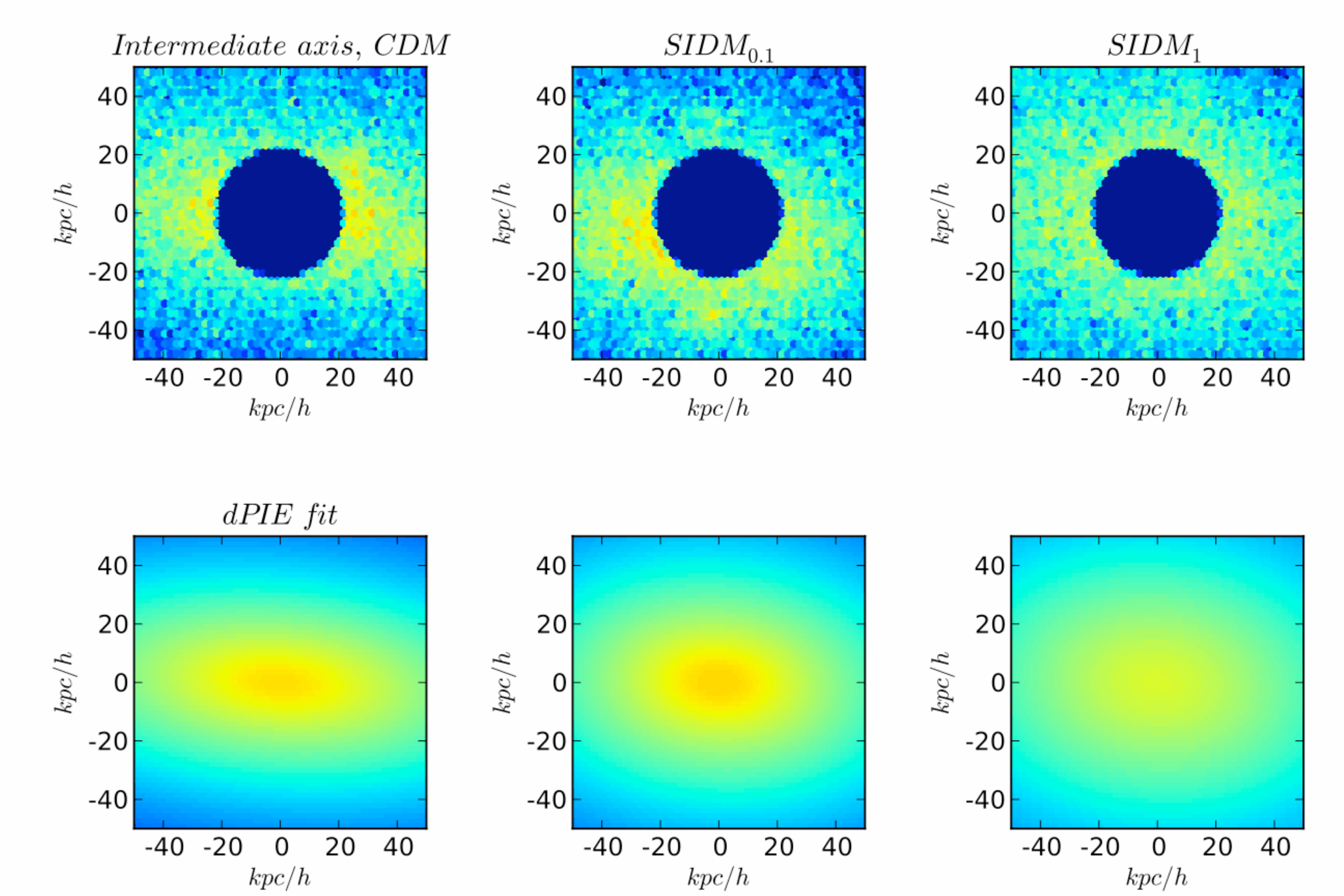}
\end{center}
\caption{\label{fig:dPIE_s1} Same as Fig. \ref{fig:dPIE_CDM} but with the line-of-sight along the intermediate axis of the halo.  The ellipticities here are: CDM, $e=0.59$; $\sigma/m = 0.1\hbox{ cm}^2/\hbox{g}$, $e=0.38$; $\sigma/m=1\hbox{ cm}^2/\hbox{g}$, $e=0.32$. }
\end{figure*}

Examples of the dPIE fits are shown in Figs. \ref{fig:dPIE_CDM} and \ref{fig:dPIE_s1}, in which we show the surface densities of one massive halo ($M_{\mathrm{vir}} = 1.8\times 10^{14}M_\odot$) along the major and intermediate principal axes of the halo moment-of-inertia tensor as well as the best-fit dPIE surface densities.  The central regions of the halos are masked for $R < r_{\mathrm{min}}$, the region where the shape profiles in three dimensions are not converged.  The halo appears rounder and more dense if projected along the major axis rather than the intermediate axis, just as we saw for another halo in Figs. \ref{fig:major4} and \ref{fig:intermed4}.  While there are noticeable differences between the CDM and $\sigma/m = 1\hbox{ cm}^2/\hbox{ g}$ surface densities at small projected radii, the differences between CDM and $\sigma/m = 0.1\hbox{ cm}^2/\hbox{g}$ are more subtle.

\begin{figure*}
\begin{center}
\includegraphics[width=0.9\textwidth]{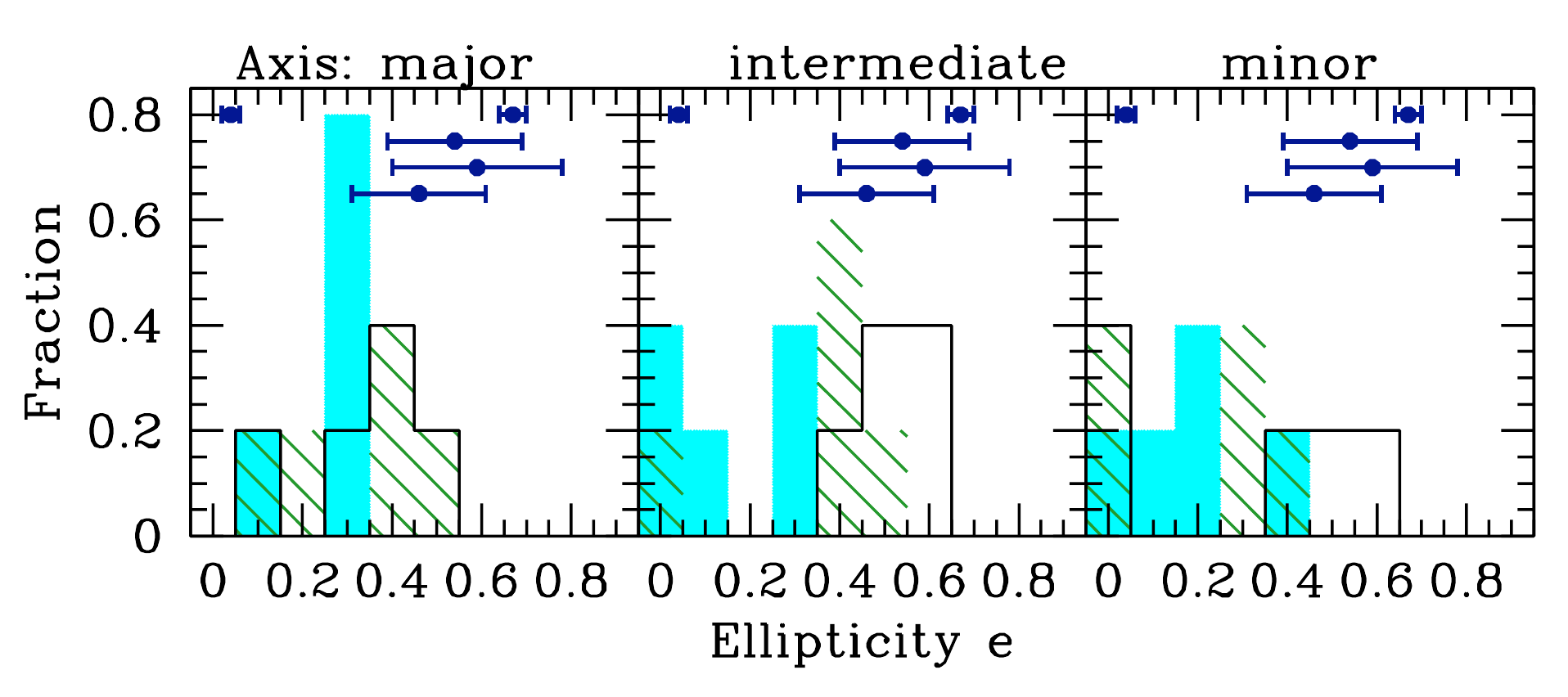}
\end{center}
\caption{\label{fig:locuss} Ellipticity $e$ (Eq. \ref{eq:e}) of halos fit with dPIE profiles for three different projections.  The solid black histograms show $e$ values for the five most massive CDM halos in the 50 $h^{-1}$Mpc simulation, the cyan histogram shows the same halos in the SIDM$_{1}$ simulation, and the green hatched histogram is for SIDM$_{0.1}$.  The dark blue points with uncertainties show the best-fit ellipticities and their 1-$\sigma$ uncertainties from dPIE modeling of the five LoCuSS clusters with $\sigma_0$ and $r_{core}$ similar to those of the simulated clusters \citep{richard2010}.}
\end{figure*}

To make a quantitative comparison between our simulations and the LoCuSS observations, we examine only those LoCuSS clusters that have $\sigma_0$ and $r_{core}$ in a similar range as the fits to the five most massive halos in our simulations.  This restricts the number of relevant LoCuSS cluster to five.  In Fig. \ref{fig:locuss}, we show the ellipticity $e$ distribution of the five most massive clusters in the CDM (black), $\sigma/m=1\hbox{ cm}^2/\hbox{g}$ (cyan), and $\sigma/m = 0.1\hbox{ cm}^2/\hbox{g}$ (green hatched) simulations for lines of sight along the three principal axes of the moment-of-inertia tensors.  The dark blue points with error bars show the central values and 1-$\sigma$ uncertainties in $e$ for the cluster halos in the lens-model fits for the five similar LoCuSS clusters.  As expected, the ellipticities are highest for the intermediate-axis line of sight.  In this instance, the ellipticities of the CDM and $\sigma/m = 0.1\hbox{ cm}^2/\hbox{g}$ halos are more consistent with the LoCuSS sample than the $\sigma/m = 1\hbox{ cm}^2/\hbox{g}$ halos.  However, the lensing probability is highest for lines of sight closely aligned to the major axis \citep{vandeven2009,mandelbaum2009}, and in this case even the CDM halos appear slightly rounder than the LoCuSS clusters.  The $\sigma/m = 1\hbox{ cm}^2/\hbox{g}$ halos are definitely too round.

Based on this initial set of LoCuSS halos, we believe it is safe to say that $\sigma/m = 0.1\hbox{ cm}^2/\hbox{g}$ is at least as consistent with observations as CDM, but that there is significant tension with $\sigma/m = 1\hbox{ cm}^2/g$.  There are several things that preclude us from making any statements stronger than this.  First, we have a small sample of both simulated and observed galaxy-cluster halos.  Second, we do not know the virial mass or alignment of the LoCuSS galaxy clusters, nor do we have a good handle on the selection function of the survey.  This means that we cannot make a direct comparison between the simulations and the observations.  Third, we do not have any galaxy-cluster-mass halos with $M_{\mathrm{vir}} > 2.2\times 10^{14}M_\odot$ in our simulations, and thus our ellipticity probability distributions for the lowest-$\sigma_0$ LoCuSS clusters are almost certainly biased, although it is not clear in which direction that bias goes.  Based on the X-ray data, it appears likely that several of the LoCuSS clusters in our subsample are more massive than any simulated cluster even though $\sigma_0$ and $r_{core}$ are similar to the simulated clusters \citep{richard2010}.  Given that the dPIE fits are made based on the very central regions of the clusters, it is not unexpected that virial masses of the clusters can be very different even if the inner regions look similar.  Finally, we do not include baryons in our simulations; it remains unclear how the presence of baryons alters the density profile and shape of galaxy clusters \citep{scannapieco2011}.

However, it is fair to say that this ensemble of observed galaxy clusters already places stronger constraints on the SIDM cross section than the other constraints we considered in this section, in particular the constraint from MS 2137-23.  While $\sigma/m = 1\hbox{ cm}^2/\hbox{g}$ is easily allowed by our reanalysis of the MS 2137-23 constraint (modulo the uncertainty in the normalization of the convergence), this value of the SIDM cross section is in some tension with the LoCuSS cluster sample.  A more quantitative limit, though, will only be possible with better theoretical predictions for the shapes of cluster-mass halos and a careful analysis of observed cluster selection functions.  In the future, we will simulate larger dark-matter halos and perform mock observations of them to find a better quantitative mapping between observations and SIDM cross-section limits.

It may be difficult to probe cross sections as small as $\sigma/m = 0.1\hbox{ cm}^2/\hbox{g}$ based on cluster halo shapes alone, though.  We get a sense of how hard it may be to use strong lensing to probe halo shapes, and hence small self-interaction cross sections, in Fig. \ref{fig:dPIE_CDM}, in which the shapes of the simulated CDM and $\sigma/m=0.1\hbox{ cm}^2/\hbox{g}$ halos do not look very different on the scales to which strong lensing is sensitive.  Moreover, in Rocha et al. (2012), we estimated that core sizes for cluster-mass halos should be $\lesssim 20\hbox{ kpc}$ if $\sigma/m = 0.1\hbox{ cm}^2/\hbox{g}$.  These sizes are similar to the effective radii of the brightest cluster galaxies (BCGs) at the centers of halos.  There are several implications of this fact.   First, it means that stellar kinematics of the BGCs will be important for probing the dark-matter halo on scales for which scattering matters.  Strong lensing is less sensitive to both the density profile and halo shapes at such small scales (see, e.g., Fig. 3 in \citet{newman2011}).    Second, it means that any inferences on the dark-matter halo on such scales depends on careful and accurate modeling of the BGCs in the data analysis.  Third, we would have to model the behavior of SIDM in the presence of a significant baryon-generated gravitational potential, and to explore the coevolution of dark-matter halos and BGCs.  In particular, simulations of isolated disky galaxies indicate that the presence of baryons tends to make the dark-matter distribution more spherical, although it is not clear how much the dark-matter distribution changes if the central galaxy is elliptical instead \citep{debattista2008}.  We note that while these issues also have implications for SIDM constraints based on radial density profiles or central densities, they may be more serious for the shape-based constraints because the shapes of $\sigma/m = 0.1\hbox{ cm}^2/\hbox{g}$ are already so similar to CDM halos for small radii even in the absence of baryons.

\section{Conclusions}\label{sec:conclusion}
The takeaway message of this work is that mapping observations to constraints on the self-interaction cross section of dark matter is significantly more subtle than previously assumed, and as such, constraints based on halo shapes are, at present, one to two orders of magnitude weaker than previously claimed.

There are three primary reasons that contribute to this conclusion.  First, the observational probes (gravitational lensing and X-ray surface brightness) of halo shapes are actually probes of some moment of the mass distribution.  For lensing, the observational probes are also sensitive to all material along the line of sight.  While SIDM makes the three-dimensional density distribution significantly rounder within some inner radius $r$, the surface density will in general not be spherically symmetric at a projected radius $R=r$.  The surface densities are affected by material well outside the core set by scatterings where material is still quite triaxial (Fig. \ref{fig:3drrvir}). Previous constraints were made under the assumption that the observations tracked the three-dimensional halo shape for fixed projected radius.  This is a less troublesome assumption for X-ray isophotes, since it is weighted by the square of the gas distribution, and hence sensitive to the central regions. The shapes measured should be related most closely to the shapes of the enclosed mass profile. So, to probe cross-sections as small as $\sigma/m = 0.1  \hbox{ cm}^2/\hbox{g}$, one needs to get down to ${\cal O}(10\, \rm kpc)$ from the center of the halos. The contribution of stars in this region makes it difficult to robustly estimate the shape of the dark matter profile and it also makes it difficult to get a large ensemble of galaxies for this study. 

Second, there is a fair bit of scatter added by assembly history to the observed shapes and the scatter is large enough that it precludes using a small number of objects to set constraints on SIDM cross sections. Finally, although we find that the three-dimensional shape of halos begins to become more spherical than CDM at radii where the local interaction rate is fairly low, $\Gamma(r) \approx 0.1 \, H_0$, there is a fair amount of triaxiality even when $\Gamma(r) \approx H_0$, a fact that was not appreciated in earlier studies \citep[e.g.,][]{miralda2002,feng2010d}  

We find that the convergence map of MS 2137-23 ($M_{\mathrm{vir}} \sim 10^{15}M_\odot$) allows a velocity-independent SIDM cross section of $\sigma/m = 1\hbox{ cm}^2/\hbox{g}$. The X-ray isophotes of NGC 720 ($M_{\mathrm{vir}} \sim 10^{13} M_\odot$) likely rule out $\sigma/m = 1\hbox{ cm}^2/\hbox{g}$, but are consistent with $\sigma/m = 0.1\hbox{ cm}^2/\hbox{g}$ at radii where we can resolve shapes in our simulations.  Based on a preliminary comparison to lensing models of LoCuSS clusters, we conclude that $\sigma/m = 0.1\hbox{ cm}^2/\hbox{g}$ is as consistent with observations as CDM but that 
 $\sigma/m = 1\hbox{ cm}^2/\hbox{g}$ is likely too large to be consistent with the observed shapes of those clusters.  

Cross sections in this range are very interesting.    In Rocha et al. (2012), we show that a cross section in the neighborhood of $\sigma/m = 0.1 \hbox{ cm}^2/\hbox{g}$ could solve the ``Too Big to Fail'' problem for the Milky Way dwarf spheroidals \citep{boylan-kolchin2011b}, the core-cusp problem in LSB galaxies \citep{kuzio2010,deblok2010}, as well as the shallow density profiles of the galaxy clusters in \citet{sand2008,newman2009}, and \citet{newman2011} while not undershooting their central densities or overshooting the core sizes.  Cross sections in this range are also consistent with other density-profile-based and subhalo-based constraints \citep{yoshida:weak:2000,gnedin2001}.

Since the current set of observations appear to be consistent with a SIDM cross section of $\sigma/m = 0.1\hbox{ cm}^2/\hbox{g}$, there are two relevant questions for shape-based SIDM constraints.  Will shape-based constraints be competitive with other types of SIDM constraints?  And 
what will it take to get down to $\sigma/m \sim 0.1\hbox{ cm}^2/\hbox{g}$ with shapes? 

Upon closer inspection, our view is that constraints using existing data could be pushed below $\sigma/m = 1\hbox{ cm}^2/\hbox{g}$, but it is not yet clear that we can get to $\sigma/m \sim 0.1\hbox{ cm}^2/\hbox{g}$.  While X-ray isophotes of the elliptical galaxy NGC 720 are consistent with $\sigma/m = 0.1\hbox{ cm}^2/\hbox{g}$, there are some differences and a larger ensemble of elliptical galaxies may be able to test that.  There are a number of other elliptical galaxies for which high-resolution X-ray data exist \citep[e.g.,][]{humphrey2006}, but they lack the detailed shape measurements of NGC 720. So better constraints could result from X-ray shape analysis for these galaxies.  For clusters, based on our quick pass through the LoCuSS results, lensing-based shape constraints on SIDM could also extend well below $\sigma/m = 1\hbox{ cm}^2/\hbox{g}$ if simulations are performed of a statistically significant number of massive galaxy clusters.
However, in studies of both galaxies and clusters, it is likely that the measured densities in the inner regions would be a better way to test for signatures of self-interacting dark matter.

\section*{Acknowledgments}
AHGP is supported by a Gary McCue Fellowship through the Center for Cosmology at UC Irvine, NASA Grant No. NNX09AD09G and NSF grant 0855462.  MR was supported by a CONACYT doctoral Fellowship and NASA grant NNX09AG01G.  
JSB was partially supported by the Miller Institute for Basic Research in Science during a Visiting Miller Professorship in the Department of Astronomy at the University of California Berkeley. This research was supported in part by the Perimeter Institute of Theoretical Physics during a visit by MK. Research at Perimeter Institute is supported by the Government of Canada through Industry Canada and by the Province of Ontario through the Ministry of Economic Development and Innovation. 
This research was also supported in part by the National Science Foundation under Grant No. NSF PHY11-25915 during a visit by AHGP to the Kavli Institute for Theoretical Physics.  AHGP thanks Michael Boylan-Kolchin for the use of his three-dimensional shape code.


\begin{thebibliography}{110}
\expandafter\ifx\csname natexlab\endcsname\relax\def\natexlab#1{#1}\fi

\bibitem[{{Ackermann} {et~al}\mbox{.}(2011){Ackermann}, {Ajello}, {Albert},
  {Atwood}, {Baldini}, {Ballet}, {Barbiellini}, {Bastieri}, {Bechtol},
  {Bellazzini}, {Berenji}, {Blandford}, {Bloom}, {Bonamente}, {Borgland},
  {Bregeon}, {Brigida}, {Bruel}, {Buehler}, {Burnett}, {Buson}, {Caliandro},
  {Cameron}, {Ca{\~n}adas}, {Caraveo}, {Casandjian}, {Cecchi}, {Charles},
  {Chekhtman}, {Chiang}, {Ciprini}, {Claus}, {Cohen-Tanugi}, {Conrad},
  {Cutini}, {de Angelis}, {de Palma}, {Dermer}, {Digel}, {Do Couto E Silva},
  {Drell}, {Drlica-Wagner}, {Falletti}, {Favuzzi}, {Fegan}, {Ferrara},
  {Fukazawa}, {Funk}, {Fusco}, {Gargano}, {Gasparrini}, {Gehrels}, {Germani},
  {Giglietto}, {Giordano}, {Giroletti}, {Glanzman}, {Godfrey}, {Grenier},
  {Guiriec}, {Gustafsson}, {Hadasch}, {Hayashida}, {Hays}, {Hughes}, {Jeltema},
  {J{\'o}hannesson}, {Johnson}, {Johnson}, {Kamae}, {Katagiri}, {Kataoka},
  {Kn{\"o}dlseder}, {Kuss}, {Lande}, {Latronico}, {Lionetto}, {Llena Garde},
  {Longo}, {Loparco}, {Lott}, {Lovellette}, {Lubrano}, {Madejski}, {Mazziotta},
  {McEnery}, {Mehault}, {Michelson}, {Mitthumsiri}, {Mizuno}, {Monte},
  {Monzani}, {Morselli}, {Moskalenko}, {Murgia}, {Naumann-Godo}, {Norris},
  {Nuss}, {Ohsugi}, {Okumura}, {Omodei}, {Orlando}, {Ormes}, {Ozaki},
  {Paneque}, {Parent}, {Pesce-Rollins}, {Pierbattista}, {Piron}, {Pivato},
  {Porter}, {Profumo}, {Rain{\`o}}, {Razzano}, {Reimer}, {Reimer}, {Ritz},
  {Roth}, {Sadrozinski}, {Sbarra}, {Scargle}, {Schalk}, {Sgr{\`o}}, {Siskind},
  {Spandre}, {Spinelli}, {Strigari}, {Suson}, {Tajima}, {Takahashi}, {Tanaka},
  {Thayer}, {Thayer}, {Thompson}, {Tibaldo}, {Tinivella}, {Torres}, {Troja},
  {Uchiyama}, {Vandenbroucke}, {Vasileiou}, {Vianello}, {Vitale}, {Waite},
  {Wang}, {Winer}, {Wood}, {Wood}, {Yang}, {Zimmer}, {Kaplinghat}, \&
  {Martinez}}]{ackermann2011a}
{Ackermann} M. {et~al.}, 2011, Phys. Rev. Lett., 107, 241302

\bibitem[{{Allgood} {et~al}\mbox{.}(2006){Allgood}, {Flores}, {Primack},
  {Kravtsov}, {Wechsler}, {Faltenbacher}, \& {Bullock}}]{allgood2006}
{Allgood} B., {Flores} R.~A., {Primack} J.~R., {Kravtsov} A.~V., {Wechsler}
  R.~H., {Faltenbacher} A., {Bullock} J.~S., 2006, \mnras, 367, 1781

\bibitem[{{Arkani-Hamed} {et~al}\mbox{.}(2009){Arkani-Hamed}, {Finkbeiner},
  {Slatyer}, \& {Weiner}}]{arkanihamed2009}
{Arkani-Hamed} N., {Finkbeiner} D.~P., {Slatyer} T.~R., {Weiner} N., 2009,
  \prd, 79, 015014

\bibitem[{{Atlas Collaboration}(2012)}]{atlas2012}
{Atlas Collaboration}, 2012, Phys. Lett. B, 709, 137

\bibitem[{{Baer}, {Barger} \& {Mustafayev}(2012){Baer}, {Barger}, \&
  {Mustafayev}}]{baer2012}
{Baer} H., {Barger} V., {Mustafayev} A., 2012, JHEP, 5, 91

\bibitem[{{Baudis}(2012)}]{baudis2012}
{Baudis} L., 2012, ArXiv e-prints, 1201.2402

\bibitem[{{Bertone} {et~al}\mbox{.}(2012){Bertone}, {Cumberbatch}, {Ruiz de
  Austri}, \& {Trotta}}]{bertone2012}
{Bertone} G., {Cumberbatch} D., {Ruiz de Austri} R., {Trotta} R., 2012, JCAP,
  1, 4

\bibitem[{{Binney} \& {Strimpel}(1978)}]{binney1978}
{Binney} J., {Strimpel} O., 1978, \mnras, 185, 473

\bibitem[{{Binney} \& {Tremaine}(2008)}]{binney2008}
{Binney} J., {Tremaine} S., 2008, {Galactic Dynamics}. Princeton, NJ, Princeton
  University Press

\bibitem[{{B{\"o}hringer} {et~al}\mbox{.}(2004){B{\"o}hringer}, {Schuecker},
  {Guzzo}, {Collins}, {Voges}, {Cruddace}, {Ortiz-Gil}, {Chincarini}, {De
  Grandi}, {Edge}, {MacGillivray}, {Neumann}, {Schindler}, \&
  {Shaver}}]{boehringer2004}
{B{\"o}hringer} H. {et~al.}, 2004, \aap, 425, 367

\bibitem[{{Boylan-Kolchin}, {Bullock} \& {Kaplinghat}(2011){Boylan-Kolchin},
  {Bullock}, \& {Kaplinghat}}]{boylan-kolchin2011}
{Boylan-Kolchin} M., {Bullock} J.~S., {Kaplinghat} M., 2011, \mnras, 415, L40

\bibitem[{{Boylan-Kolchin}, {Bullock} \& {Kaplinghat}(2012){Boylan-Kolchin},
  {Bullock}, \& {Kaplinghat}}]{boylan-kolchin2011b}
{Boylan-Kolchin} M., {Bullock} J.~S., {Kaplinghat} M., 2012, \mnras, 422, 1203

\bibitem[{{Bryan} \& {Norman}(1998)}]{bryan1998}
{Bryan} G.~L., {Norman} M.~L., 1998, \apj, 495, 80

\bibitem[{{Buckley} \& {Fox}(2010)}]{buckley2010b}
{Buckley} M.~R., {Fox} P.~J., 2010, \prd, 81, 083522

\bibitem[{{Buote} \& {Canizares}(1994)}]{buote1994}
{Buote} D.~A., {Canizares} C.~R., 1994, \apj, 427, 86

\bibitem[{{Buote} \& {Canizares}(1996)}]{buote1996}
{Buote} D.~A., {Canizares} C.~R., 1996, \apj, 468, 184

\bibitem[{{Buote} \& {Canizares}(1998{\natexlab{a}})}]{buote1998b}
{Buote} D.~A., {Canizares} C.~R., 1998{\natexlab{a}}, in Astronomical Society
  of the Pacific Conference Series, Vol. 136, Galactic Halos, {D.~Zaritsky},
  ed., p. 289

\bibitem[{{Buote} \& {Canizares}(1998{\natexlab{b}})}]{buote1998}
{Buote} D.~A., {Canizares} C.~R., 1998{\natexlab{b}}, \mnras, 298, 811

\bibitem[{{Buote} {et~al}\mbox{.}(2002){Buote}, {Jeltema}, {Canizares}, \&
  {Garmire}}]{buote2002}
{Buote} D.~A., {Jeltema} T.~E., {Canizares} C.~R., {Garmire} G.~P., 2002, \apj,
  577, 183

\bibitem[{{Burkert}(1995)}]{burkert1995}
{Burkert} A., 1995, \apjl, 447, L25

\bibitem[{{Cohen} {et~al}\mbox{.}(2010){Cohen}, {Phalen}, {Pierce}, \&
  {Zurek}}]{cohen2010}
{Cohen} T., {Phalen} D.~J., {Pierce} A., {Zurek} K.~M., 2010, \prd, 82, 056001

\bibitem[{{Col{\'{\i}}n} {et~al}\mbox{.}(2002){Col{\'{\i}}n}, {Avila-Reese},
  {Valenzuela}, \& {Firmani}}]{colin2002}
{Col{\'{\i}}n} P., {Avila-Reese} V., {Valenzuela} O., {Firmani} C., 2002, \apj,
  581, 777

\bibitem[{{Cotta} {et~al}\mbox{.}(2012){Cotta}, {Drlica-Wagner}, {Murgia},
  {Bloom}, {Hewett}, \& {Rizzo}}]{cotta2011}
{Cotta} R.~C., {Drlica-Wagner} A., {Murgia} S., {Bloom} E.~D., {Hewett} J.~L.,
  {Rizzo} T.~G., 2012, JCAP, 4, 16

\bibitem[{{Dalcanton} \& {Hogan}(2001)}]{dalcanton2001}
{Dalcanton} J.~J., {Hogan} C.~J., 2001, \apj, 561, 35

\bibitem[{{Dav{\'e}} {et~al}\mbox{.}(2001){Dav{\'e}}, {Spergel}, {Steinhardt},
  \& {Wandelt}}]{dave2001}
{Dav{\'e}} R., {Spergel} D.~N., {Steinhardt} P.~J., {Wandelt} B.~D., 2001,
  \apj, 547, 574

\bibitem[{{de Blok}(2010)}]{deblok2010}
{de Blok} W.~J.~G., 2010, Advances in Astronomy, 2010

\bibitem[{{de Blok} {et~al}\mbox{.}(2008){de Blok}, {Walter}, {Brinks},
  {Trachternach}, {Oh}, \& {Kennicutt}}]{deblok2008}
{de Blok} W.~J.~G., {Walter} F., {Brinks} E., {Trachternach} C., {Oh} S.-H.,
  {Kennicutt} R.~C., 2008, Astron. J., 136, 2648

\bibitem[{{Debattista} {et~al}\mbox{.}(2008){Debattista}, {Moore}, {Quinn},
  {Kazantzidis}, {Maas}, {Mayer}, {Read}, \& {Stadel}}]{debattista2008}
{Debattista} V.~P., {Moore} B., {Quinn} T., {Kazantzidis} S., {Maas} R.,
  {Mayer} L., {Read} J., {Stadel} J., 2008, \apj, 681, 1076

\bibitem[{{Diemand} {et~al}\mbox{.}(2008){Diemand}, {Kuhlen}, {Madau}, {Zemp},
  {Moore}, {Potter}, \& {Stadel}}]{diemand2008}
{Diemand} J., {Kuhlen} M., {Madau} P., {Zemp} M., {Moore} B., {Potter} D.,
  {Stadel} J., 2008, \nat, 454, 735

\bibitem[{{Dobler} \& {Keeton}(2006)}]{dobler2006}
{Dobler} G., {Keeton} C.~R., 2006, Mon. Not. R. Astron. Soc., 365, 1243

\bibitem[{{Dubinski} \& {Carlberg}(1991)}]{dubinski1991}
{Dubinski} J., {Carlberg} R.~G., 1991, \apj, 378, 496

\bibitem[{{Dutton} {et~al}\mbox{.}(2011){Dutton}, {Conroy}, {van den Bosch},
  {Simard}, {Mendel}, {Courteau}, {Dekel}, {More}, \& {Prada}}]{dutton2011}
{Dutton} A.~A. {et~al.}, 2011, \mnras, 416, 322

\bibitem[{{Ebeling} {et~al}\mbox{.}(2000){Ebeling}, {Edge}, {Allen},
  {Crawford}, {Fabian}, \& {Huchra}}]{ebeling2000}
{Ebeling} H., {Edge} A.~C., {Allen} S.~W., {Crawford} C.~S., {Fabian} A.~C.,
  {Huchra} J.~P., 2000, \mnras, 318, 333

\bibitem[{{Fabricant}, {Rybicki} \& {Gorenstein}(1984){Fabricant}, {Rybicki},
  \& {Gorenstein}}]{fabricant1984}
{Fabricant} D., {Rybicki} G., {Gorenstein} P., 1984, \apj, 286, 186

\bibitem[{{Feldman}, {Kors} \& {Nath}(2007){Feldman}, {Kors}, \&
  {Nath}}]{feldman2007}
{Feldman} D., {Kors} B., {Nath} P., 2007, \prd, 75, 023503

\bibitem[{{Feng} {et~al}\mbox{.}(2009){Feng}, {Kaplinghat}, {Tu}, \&
  {Yu}}]{feng2009}
{Feng} J.~L., {Kaplinghat} M., {Tu} H., {Yu} H.-B., 2009, JCAP, 7, 4

\bibitem[{{Feng}, {Kaplinghat} \& {Yu}(2010){Feng}, {Kaplinghat}, \&
  {Yu}}]{feng2010d}
{Feng} J.~L., {Kaplinghat} M., {Yu} H.-B., 2010, Phys. Rev. Lett., 104, 151301

\bibitem[{{Feng} \& {Kumar}(2008)}]{feng2008}
{Feng} J.~L., {Kumar} J., 2008, Phys. Rev. Lett., 101, 231301

\bibitem[{{Feng}, {Rentala} \& {Surujon}(2012){Feng}, {Rentala}, \&
  {Surujon}}]{feng2012}
{Feng} J.~L., {Rentala} V., {Surujon} Z., 2012, \prd, 85, 055003

\bibitem[{{Foot}(2007)}]{foot2007}
{Foot} R., 2007, International Journal of Modern Physics A, 22, 4951

\bibitem[{{Fort} {et~al}\mbox{.}(1992){Fort}, {Le Fevre}, {Hammer}, \&
  {Cailloux}}]{fort1992}
{Fort} B., {Le Fevre} O., {Hammer} F., {Cailloux} M., 1992, \apjl, 399, L125

\bibitem[{{Fox} {et~al}\mbox{.}(2011){Fox}, {Harnik}, {Kopp}, \&
  {Tsai}}]{fox2011b}
{Fox} P.~J., {Harnik} R., {Kopp} J., {Tsai} Y., 2011, ArXiv e-prints, 1109.4398

\bibitem[{{Fox} \& {Poppitz}(2009)}]{fox2009}
{Fox} P.~J., {Poppitz} E., 2009, \prd, 79, 083528

\bibitem[{{Gavazzi}(2005)}]{gavazzi2005}
{Gavazzi} R., 2005, \aap, 443, 793

\bibitem[{{Gavazzi} {et~al}\mbox{.}(2003){Gavazzi}, {Fort}, {Mellier},
  {Pell{\'o}}, \& {Dantel-Fort}}]{gavazzi2003}
{Gavazzi} R., {Fort} B., {Mellier} Y., {Pell{\'o}} R., {Dantel-Fort} M., 2003,
  \aap, 403, 11

\bibitem[{{Gavazzi} {et~al}\mbox{.}(2007){Gavazzi}, {Treu}, {Rhodes},
  {Koopmans}, {Bolton}, {Burles}, {Massey}, \& {Moustakas}}]{gavazzi2007}
{Gavazzi} R., {Treu} T., {Rhodes} J.~D., {Koopmans} L.~V.~E., {Bolton} A.~S.,
  {Burles} S., {Massey} R.~J., {Moustakas} L.~A., 2007, \apj, 667, 176

\bibitem[{{Gentile} {et~al}\mbox{.}(2007){Gentile}, {Salucci}, {Klein}, \&
  {Granato}}]{gentile2007}
{Gentile} G., {Salucci} P., {Klein} U., {Granato} G.~L., 2007, \mnras, 375, 199

\bibitem[{{Geringer-Sameth} \& {Koushiappas}(2011)}]{geringer-sameth2011}
{Geringer-Sameth} A., {Koushiappas} S.~M., 2011, Phys. Rev. Lett., 107, 241303

\bibitem[{{Gill} {et~al}\mbox{.}(2009){Gill}, {Young}, {Draskovic},
  {Honscheid}, {Lin}, {Kuropatkin}, {Martini}, {Peeples}, {Rozo}, {Smith}, \&
  {Weinberg}}]{gill2009}
{Gill} M.~S.~S. {et~al.}, 2009, ArXiv e-prints, 0909.3856

\bibitem[{{Gnedin} \& {Ostriker}(2001)}]{gnedin2001}
{Gnedin} O.~Y., {Ostriker} J.~P., 2001, \apj, 561, 61

\bibitem[{{Griest}(1988)}]{griest1988b}
{Griest} K., 1988, \prd, 38, 2357

\bibitem[{{Hahn} \& {Abel}(2011)}]{hahn2011}
{Hahn} O., {Abel} T., 2011, \mnras, 415, 2101

\bibitem[{{Hennawi} \& {Ostriker}(2002)}]{hennawi2002}
{Hennawi} J.~F., {Ostriker} J.~P., 2002, \apj, 572, 41

\bibitem[{{Hernquist}(1990)}]{hernquist1990}
{Hernquist} L., 1990, \apj, 356, 359

\bibitem[{{Hogan} \& {Dalcanton}(2000)}]{hogan2000}
{Hogan} C.~J., {Dalcanton} J.~J., 2000, \prd, 62, 063511

\bibitem[{{Humphrey} {et~al}\mbox{.}(2006){Humphrey}, {Buote}, {Gastaldello},
  {Zappacosta}, {Bullock}, {Brighenti}, \& {Mathews}}]{humphrey2006}
{Humphrey} P.~J., {Buote} D.~A., {Gastaldello} F., {Zappacosta} L., {Bullock}
  J.~S., {Brighenti} F., {Mathews} W.~G., 2006, \apj, 646, 899

\bibitem[{{Ibe} \& {Yu}(2010)}]{ibe2010}
{Ibe} M., {Yu} H.-B., 2010, Phys. Lett. B, 692, 70

\bibitem[{{Jungman}, {Kamionkowski} \& {Griest}(1996){Jungman}, {Kamionkowski},
  \& {Griest}}]{jungman1996}
{Jungman} G., {Kamionkowski} M., {Griest} K., 1996, Phys. Rep., 267, 195

\bibitem[{{Khlopov}, {Stephan} \& {Fargion}(2006){Khlopov}, {Stephan}, \&
  {Fargion}}]{khlopov2006}
{Khlopov} M.~Y., {Stephan} C.~A., {Fargion} D., 2006, Classical and Quantum
  Gravity, 23, 7305

\bibitem[{{Kneib} \& {Natarajan}(2011)}]{kneib2011}
{Kneib} J.-P., {Natarajan} P., 2011, \aapr, 19, 47

\bibitem[{{Knollmann} \& {Knebe}(2009)}]{knollmann2009}
{Knollmann} S.~R., {Knebe} A., 2009, \apjs, 182, 608

\bibitem[{{Koay}(2012)}]{koay2012}
{Koay} S.~A., 2012, in European Physical Journal Web of Conferences, Vol.~28,
  European Physical Journal Web of Conferences, p. 9008

\bibitem[{{Kochanek} \& {White}(2000)}]{kochanek2000}
{Kochanek} C.~S., {White} M., 2000, \apj, 543, 514

\bibitem[{{Komatsu} {et~al}\mbox{.}(2011){Komatsu}, {Smith}, {Dunkley},
  {Bennett}, {Gold}, {Hinshaw}, {Jarosik}, {Larson}, {Nolta}, {Page},
  {Spergel}, {Halpern}, {Hill}, {Kogut}, {Limon}, {Meyer}, {Odegard}, {Tucker},
  {Weiland}, {Wollack}, \& {Wright}}]{komatsu2011}
{Komatsu} E. {et~al.}, 2011, \apjs, 192, 18

\bibitem[{{Kuzio de Naray} {et~al}\mbox{.}(2010){Kuzio de Naray}, {Martinez},
  {Bullock}, \& {Kaplinghat}}]{kuzio2010}
{Kuzio de Naray} R., {Martinez} G.~D., {Bullock} J.~S., {Kaplinghat} M., 2010,
  \apj, 710, L161

\bibitem[{{Kuzio de Naray}, {McGaugh} \& {de Blok}(2008){Kuzio de Naray},
  {McGaugh}, \& {de Blok}}]{kuzio2008}
{Kuzio de Naray} R., {McGaugh} S.~S., {de Blok} W.~J.~G., 2008, \apj, 676, 920

\bibitem[{{Loeb} \& {Weiner}(2011)}]{loeb2011}
{Loeb} A., {Weiner} N., 2011, Phys. Rev. Lett., 106, 171302

\bibitem[{{LSST Science Collaborations}(2009)}]{lsst2009}
{LSST Science Collaborations}, 2009, ArXiv e-prints, 0912.0201

\bibitem[{{Mandelbaum}, {van de Ven} \& {Keeton}(2009){Mandelbaum}, {van de
  Ven}, \& {Keeton}}]{mandelbaum2009}
{Mandelbaum} R., {van de Ven} G., {Keeton} C.~R., 2009, \mnras, 398, 635

\bibitem[{{Marriage} {et~al}\mbox{.}(2011){Marriage}, {Acquaviva}, {Ade},
  {Aguirre}, {Amiri}, {Appel}, {Barrientos}, {Battistelli}, {Bond}, {Brown},
  {Burger}, {Chervenak}, {Das}, {Devlin}, {Dicker}, {Bertrand Doriese},
  {Dunkley}, {D{\"u}nner}, {Essinger-Hileman}, {Fisher}, {Fowler}, {Hajian},
  {Halpern}, {Hasselfield}, {Hern{\'a}ndez-Monteagudo}, {Hilton}, {Hilton},
  {Hincks}, {Hlozek}, {Huffenberger}, {Handel Hughes}, {Hughes}, {Infante},
  {Irwin}, {Baptiste Juin}, {Kaul}, {Klein}, {Kosowsky}, {Lau}, {Limon}, {Lin},
  {Lupton}, {Marsden}, {Martocci}, {Mauskopf}, {Menanteau}, {Moodley},
  {Moseley}, {Netterfield}, {Niemack}, {Nolta}, {Page}, {Parker}, {Partridge},
  {Quintana}, {Reese}, {Reid}, {Sehgal}, {Sherwin}, {Sievers}, {Spergel},
  {Staggs}, {Swetz}, {Switzer}, {Thornton}, {Trac}, {Tucker}, {Warne},
  {Wilson}, {Wollack}, \& {Zhao}}]{marriage2011}
{Marriage} T.~A. {et~al.}, 2011, \apj, 737, 61

\bibitem[{{McDermott}, {Yu} \& {Zurek}(2011){McDermott}, {Yu}, \&
  {Zurek}}]{mcdermott2011}
{McDermott} S.~D., {Yu} H.-B., {Zurek} K.~M., 2011, \prd, 83, 063509

\bibitem[{{Mellier}, {Fort} \& {Kneib}(1993){Mellier}, {Fort}, \&
  {Kneib}}]{mellier1993}
{Mellier} Y., {Fort} B., {Kneib} J.-P., 1993, \apj, 407, 33

\bibitem[{{Meneghetti} {et~al}\mbox{.}(2001){Meneghetti}, {Yoshida},
  {Bartelmann}, {Moscardini}, {Springel}, {Tormen}, \&
  {White}}]{meneghetti2001}
{Meneghetti} M., {Yoshida} N., {Bartelmann} M., {Moscardini} L., {Springel} V.,
  {Tormen} G., {White} S.~D.~M., 2001, \mnras, 325, 435

\bibitem[{{Miralda-Escude}(1995)}]{miralda1995}
{Miralda-Escude} J., 1995, \apj, 438, 514

\bibitem[{{Miralda-Escud{\'e}}(2002)}]{miralda2002}
{Miralda-Escud{\'e}} J., 2002, \apj, 564, 60

\bibitem[{{Navarro}, {Frenk} \& {White}(1997){Navarro}, {Frenk}, \&
  {White}}]{navarro1997}
{Navarro} J.~F., {Frenk} C.~S., {White} S.~D.~M., 1997, \apj, 490, 493

\bibitem[{{Navarro} {et~al}\mbox{.}(2004){Navarro}, {Hayashi}, {Power},
  {Jenkins}, {Frenk}, {White}, {Springel}, {Stadel}, \& {Quinn}}]{navarro2004}
{Navarro} J.~F. {et~al.}, 2004, \mnras, 349, 1039

\bibitem[{{Navarro} {et~al}\mbox{.}(2010){Navarro}, {Ludlow}, {Springel},
  {Wang}, {Vogelsberger}, {White}, {Jenkins}, {Frenk}, \&
  {Helmi}}]{navarro2010}
{Navarro} J.~F. {et~al.}, 2010, \mnras, 402, 21

\bibitem[{{Newman} {et~al}\mbox{.}(2011){Newman}, {Treu}, {Ellis}, \&
  {Sand}}]{newman2011}
{Newman} A.~B., {Treu} T., {Ellis} R.~S., {Sand} D.~J., 2011, \apjl, 728, L39+

\bibitem[{{Newman} {et~al}\mbox{.}(2009){Newman}, {Treu}, {Ellis}, {Sand},
  {Richard}, {Marshall}, {Capak}, \& {Miyazaki}}]{newman2009}
{Newman} A.~B., {Treu} T., {Ellis} R.~S., {Sand} D.~J., {Richard} J.,
  {Marshall} P.~J., {Capak} P., {Miyazaki} S., 2009, \apj, 706, 1078

\bibitem[{{Papastergis} {et~al}\mbox{.}(2011){Papastergis}, {Martin},
  {Giovanelli}, \& {Haynes}}]{papastergis2011}
{Papastergis} E., {Martin} A.~M., {Giovanelli} R., {Haynes} M.~P., 2011, \apj,
  739, 38

\bibitem[{{Pillepich}, {Porciani} \& {Reiprich}(2012){Pillepich}, {Porciani},
  \& {Reiprich}}]{pillepich2011}
{Pillepich} A., {Porciani} C., {Reiprich} T.~H., 2012, \mnras, 422, 44

\bibitem[{{Plagge} {et~al}\mbox{.}(2010){Plagge}, {Benson}, {Ade}, {Aird},
  {Bleem}, {Carlstrom}, {Chang}, {Cho}, {Crawford}, {Crites}, {de Haan},
  {Dobbs}, {George}, {Hall}, {Halverson}, {Holder}, {Holzapfel}, {Hrubes},
  {Joy}, {Keisler}, {Knox}, {Lee}, {Leitch}, {Lueker}, {Marrone}, {McMahon},
  {Mehl}, {Meyer}, {Mohr}, {Montroy}, {Padin}, {Pryke}, {Reichardt}, {Ruhl},
  {Schaffer}, {Shaw}, {Shirokoff}, {Spieler}, {Stalder}, {Staniszewski},
  {Stark}, {Vanderlinde}, {Vieira}, {Williamson}, \& {Zahn}}]{plagge2010}
{Plagge} T. {et~al.}, 2010, \apj, 716, 1118

\bibitem[{{Planck Collaboration} {et~al}\mbox{.}(2011{\natexlab{a}}){Planck
  Collaboration}, {Ade}, {Aghanim}, {Arnaud}, {Ashdown}, {Aumont},
  {Baccigalupi}, {Balbi}, {Banday}, {Barreiro}, \& et~al.}]{planck2011viii}
{Planck Collaboration} {et~al.}, 2011{\natexlab{a}}, \aap, 536, A8

\bibitem[{{Planck Collaboration} {et~al}\mbox{.}(2011{\natexlab{b}}){Planck
  Collaboration}, {Aghanim}, {Arnaud}, {Ashdown}, {Aumont}, {Baccigalupi},
  {Balbi}, {Banday}, {Barreiro}, {Bartelmann}, \& et~al.}]{planck2011ix}
{Planck Collaboration} {et~al.}, 2011{\natexlab{b}}, \aap, 536, A9

\bibitem[{{Planck Collaboration} {et~al}\mbox{.}(2011{\natexlab{c}}){Planck
  Collaboration}, {Aghanim}, {Arnaud}, {Ashdown}, {Aumont}, {Baccigalupi},
  {Balbi}, {Banday}, {Barreiro}, {Bartelmann}, \& et~al.}]{planck2011xii}
{Planck Collaboration} {et~al.}, 2011{\natexlab{c}}, \aap, 536, A12

\bibitem[{{Pospelov}, {Ritz} \& {Voloshin}(2008){Pospelov}, {Ritz}, \&
  {Voloshin}}]{pospelov2008}
{Pospelov} M., {Ritz} A., {Voloshin} M., 2008, Phys. Lett. B, 662, 53

\bibitem[{{Postman} {et~al}\mbox{.}(2012){Postman}, {Coe}, {Ben{\'{\i}}tez},
  {Bradley}, {Broadhurst}, {Donahue}, {Ford}, {Graur}, {Graves}, {Jouvel},
  {Koekemoer}, {Lemze}, {Medezinski}, {Molino}, {Moustakas}, {Ogaz}, {Riess},
  {Rodney}, {Rosati}, {Umetsu}, {Zheng}, {Zitrin}, {Bartelmann}, {Bouwens},
  {Czakon}, {Golwala}, {Host}, {Infante}, {Jha}, {Jimenez-Teja}, {Kelson},
  {Lahav}, {Lazkoz}, {Maoz}, {McCully}, {Melchior}, {Meneghetti}, {Merten},
  {Moustakas}, {Nonino}, {Patel}, {Reg{\"o}s}, {Sayers}, {Seitz}, \& {Van der
  Wel}}]{postman2012}
{Postman} M. {et~al.}, 2012, \apjs, 199, 25

\bibitem[{{Power} {et~al}\mbox{.}(2003){Power}, {Navarro}, {Jenkins}, {Frenk},
  {White}, {Springel}, {Stadel}, \& {Quinn}}]{power2003}
{Power} C., {Navarro} J.~F., {Jenkins} A., {Frenk} C.~S., {White} S.~D.~M.,
  {Springel} V., {Stadel} J., {Quinn} T., 2003, \mnras, 338, 14

\bibitem[{{Randall} {et~al}\mbox{.}(2008){Randall}, {Markevitch}, {Clowe},
  {Gonzalez}, \& {Brada{\v c}}}]{randall2008}
{Randall} S.~W., {Markevitch} M., {Clowe} D., {Gonzalez} A.~H., {Brada{\v c}}
  M., 2008, \apj, 679, 1173

\bibitem[{{Reid} {et~al}\mbox{.}(2010){Reid}, {Percival}, {Eisenstein},
  {Verde}, {Spergel}, {Skibba}, {Bahcall}, {Budavari}, {Frieman}, {Fukugita},
  {Gott}, {Gunn}, {Ivezi{\'c}}, {Knapp}, {Kron}, {Lupton}, {McKay}, {Meiksin},
  {Nichol}, {Pope}, {Schlegel}, {Schneider}, {Stoughton}, {Strauss}, {Szalay},
  {Tegmark}, {Vogeley}, {Weinberg}, {York}, \& {Zehavi}}]{breid2010}
{Reid} B.~A. {et~al.}, 2010, \mnras, 404, 60

\bibitem[{{Richard} {et~al}\mbox{.}(2010){Richard}, {Smith}, {Kneib}, {Ellis},
  {Sanderson}, {Pei}, {Targett}, {Sand}, {Swinbank}, {Dannerbauer}, {Mazzotta},
  {Limousin}, {Egami}, {Jullo}, {Hamilton-Morris}, \& {Moran}}]{richard2010}
{Richard} J. {et~al.}, 2010, \mnras, 404, 325

\bibitem[{{Sand} {et~al}\mbox{.}(2008){Sand}, {Treu}, {Ellis}, {Smith}, \&
  {Kneib}}]{sand2008}
{Sand} D.~J., {Treu} T., {Ellis} R.~S., {Smith} G.~P., {Kneib} J.-P., 2008,
  \apj, 674, 711

\bibitem[{{Scannapieco} {et~al}\mbox{.}(2012){Scannapieco}, {Wadepuhl},
  {Parry}, {Navarro}, {Jenkins}, {Springel}, {Teyssier}, {Carlson}, {Couchman},
  {Crain}, {Vecchia}, {Frenk}, {Kobayashi}, {Monaco}, {Murante}, {Okamoto},
  {Quinn}, {Schaye}, {Stinson}, {Theuns}, {Wadsley}, {White}, \&
  {Woods}}]{scannapieco2011}
{Scannapieco} C. {et~al.}, 2012, \mnras, 423, 1726

\bibitem[{{Schulz}, {Mandelbaum} \& {Padmanabhan}(2010){Schulz}, {Mandelbaum},
  \& {Padmanabhan}}]{schulz2010}
{Schulz} A.~E., {Mandelbaum} R., {Padmanabhan} N., 2010, \mnras, 408, 1463

\bibitem[{{Sigurdson}(2009)}]{sigurdson2009}
{Sigurdson} K., 2009, ArXiv e-prints, 0912.2346

\bibitem[{{Spergel} \& {Steinhardt}(2000)}]{spergel2000}
{Spergel} D.~N., {Steinhardt} P.~J., 2000, Phys. Rev. Lett., 84, 3760

\bibitem[{{Springel}(2005)}]{springel2005}
{Springel} V., 2005, \mnras, 364, 1105

\bibitem[{{Stadel} {et~al}\mbox{.}(2009){Stadel}, {Potter}, {Moore}, {Diemand},
  {Madau}, {Zemp}, {Kuhlen}, \& {Quilis}}]{stadel2009}
{Stadel} J., {Potter} D., {Moore} B., {Diemand} J., {Madau} P., {Zemp} M.,
  {Kuhlen} M., {Quilis} V., 2009, \mnras, 398, L21

\bibitem[{{Steigman} \& {Turner}(1985)}]{steigman1985}
{Steigman} G., {Turner} M.~S., 1985, Nucl. Phys. B, 253, 375

\bibitem[{{Tolman}(1934)}]{tolman1934}
{Tolman} R.~C., 1934, PNAS, 20, 169

\bibitem[{{van de Ven}, {Mandelbaum} \& {Keeton}(2009){van de Ven},
  {Mandelbaum}, \& {Keeton}}]{vandeven2009}
{van de Ven} G., {Mandelbaum} R., {Keeton} C.~R., 2009, \mnras, 398, 607

\bibitem[{{Vera-Ciro} {et~al}\mbox{.}(2011){Vera-Ciro}, {Sales}, {Helmi},
  {Frenk}, {Navarro}, {Springel}, {Vogelsberger}, \& {White}}]{vera-ciro2011}
{Vera-Ciro} C.~A., {Sales} L.~V., {Helmi} A., {Frenk} C.~S., {Navarro} J.~F.,
  {Springel} V., {Vogelsberger} M., {White} S.~D.~M., 2011, \mnras, 416, 1377

\bibitem[{{Viana} {et~al}\mbox{.}(2012){Viana}, {da Silva}, {Ramos}, {Liddle},
  {Lloyd-Davies}, {Romer}, {Kay}, {Collins}, {Hilton}, {Hosmer}, {Hoyle},
  {Mayers}, {Mehrtens}, {Miller}, {Sahl{\'e}n}, {Stanford}, \&
  {Stott}}]{viana2012}
{Viana} P.~T.~P. {et~al.}, 2012, \mnras, 2526

\bibitem[{{Vogelsberger} {et~al}\mbox{.}(2009){Vogelsberger}, {Helmi},
  {Springel}, {White}, {Wang}, {Frenk}, {Jenkins}, {Ludlow}, \&
  {Navarro}}]{vogelsberger2009}
{Vogelsberger} M. {et~al.}, 2009, \mnras, 395, 797

\bibitem[{{Vogelsberger}, {Zavala} \& {Loeb}(2012){Vogelsberger}, {Zavala}, \&
  {Loeb}}]{vogelsberger2012}
{Vogelsberger} M., {Zavala} J., {Loeb} A., 2012, \mnras, 423, 3740

\bibitem[{{XENON100 Collaboration} {et~al}\mbox{.}(2012){XENON100
  Collaboration}, {Aprile}, {Alfonsi}, {Arisaka}, {Arneodo}, {Balan}, {Baudis},
  {Bauermeister}, {Behrens}, {Beltrame}, {Bokeloh}, {Brown}, {Bruno}, {Budnik},
  {Cardoso}, {Chen}, {Choi}, {Cline}, {Colijn}, {Contreras}, {Cussonneau},
  {Decowski}, {Duchovni}, {Fattori}, {Ferella}, {Fulgione}, {Gao}, {Garbini},
  {Ghag}, {Giboni}, {Goetzke}, {Grignon}, {Gross}, {Hampel}, {Kaether},
  {Kettling}, {Kish}, {Lamblin}, {Landsman}, {Lang}, {Le Calloch}, {Levy},
  {Lim}, {Lin}, {Lindemann}, {Lindner}, {Lopes}, {Lung}, {Marrodan Undagoitia},
  {Massoli}, {Melgarejo Fernandez}, {Meng}, {Molinario}, {Nativ}, {Ni},
  {Oberlack}, {Orrigo}, {Pantic}, {Persiani}, {Plante}, {Priel}, {Rizzo},
  {Rosendahl}, {dos Santos}, {Sartorelli}, {Schreiner}, {Schumann}, {Scotto
  Lavina}, {Scovell}, {Selvi}, {Shagin}, {Simgen}, {Teymourian}, {Thers},
  {Vitells}, {Wang}, {Weber}, \& {Weinheimer}}]{xenon2012}
{XENON100 Collaboration} {et~al.}, 2012, ArXiv e-prints, 1207.5988

\bibitem[{{Yoshida} {et~al}\mbox{.}(2000{\natexlab{a}}){Yoshida}, {Springel},
  {White}, \& {Tormen}}]{yoshida2000}
{Yoshida} N., {Springel} V., {White} S.~D.~M., {Tormen} G., 2000{\natexlab{a}},
  \apj, 535, L103

\bibitem[{{Yoshida} {et~al}\mbox{.}(2000{\natexlab{b}}){Yoshida}, {Springel},
  {White}, \& {Tormen}}]{yoshida:weak:2000}
{Yoshida} N., {Springel} V., {White} S.~D.~M., {Tormen} G., 2000{\natexlab{b}},
  \apj, 544, L87

\bibitem[{{Zwaan}, {Meyer} \& {Staveley-Smith}(2010){Zwaan}, {Meyer}, \&
  {Staveley-Smith}}]{zwaan2010}
{Zwaan} M.~A., {Meyer} M.~J., {Staveley-Smith} L., 2010, \mnras, 403, 1969

\end{thebibliography}

\end{document}